\documentclass[letterpaper]{emulateapj}
\usepackage{wasysym} 
\usepackage{graphicx}
\usepackage{lscape}
\usepackage{subfigure}

\newcommand{\hii}{H\,{\sc ii}}
\newcommand{\siv}{S\,{\sc iv}}

\newcommand{\neii}{Ne\,{\sc ii}}
\newcommand{\neiii}{Ne\,{\sc iii}}
\newcommand{\sii}{S\,{\sc ii}}
\newcommand{\siii}{S\,{\sc iii}}
\newcommand{\silii}{Si\,{\sc ii}}
\newcommand{\arii}{Ar\,{\sc ii}}
\newcommand{\ariii}{Ar\,{\sc iii}}

\newcommand{\mum}{\ensuremath{\mu \rm{m}}}

\begin{document}

\title{Ongoing Massive Star Formation in NGC~604} 

\shorttitle{Ongoing Massive Star Formation in NGC~604}

\author{J.~R. Mart\'{\i}nez-Galarza\altaffilmark{1,2}, D. Hunter \altaffilmark{3}, B. Groves\altaffilmark{2,4}, B. Brandl\altaffilmark{2}}

\altaffiltext{1}{Harvard-Smithsonian Center for Astrophysics, 60 Garden Street, Cambridge, MA 02138, USA, jmartine@cfa.harvard.edu}
\altaffiltext{2}{Leiden Observatory, Leiden University, P.O. Box 9513, 2300 CA Leiden, The Netherlands}
\altaffiltext{3}{Lowell Observatory, 1400 W Mars Hill Road, 86001 AZ, USA}
\altaffiltext{4}{Max-Planck-Institut f\"ur Astronomie, K\"onigstuhl 17, D-69117 Heidelberg, Germany}

\begin{abstract}

NGC~604 is the second most massive \hii\ region in the Local Group, thus an important laboratory for massive star formation. Using a combination of observational and analytical tools that include \emph{Spitzer} spectroscopy, \emph{Herschel} photometry, \emph{Chandra} imaging, and Bayesian Spectral Energy Distribution fitting, we investigate the physical conditions in NGC\, 604, and quantify the amount of massive star formation currently taking place. We derive an average age of $4\pm 1\: $Myr and a total stellar mass of $1.6^{+1.6}_{-1.0}\times 10^5\: \rm{M}_{\astrosun}$ for the entire region, in agreement with previous optical studies. Across the region we find an effect of the X-ray field on both the abundance of aromatic molecules and the [\silii] emission. Within NGC~604 we identify several individual bright infrared sources with diameters of about $15\: $pc and luminosity weighted masses between $10^3\: \rm{M}_{\astrosun}$ and $10^4\: \rm{M}_{\astrosun}$. Their spectral properties indicate that some of these sources are embedded clusters in process of formation, which together account for $\sim$8\% of the total stellar mass in the NGC\,604 system. The variations of the radiation field strength across NGC~604 are consistent with a sequential star formation scenario, with at least two bursts in the last few million years. Our results indicate that massive star formation in NGC~604 is still ongoing, likely triggered by the earlier bursts.

\end{abstract}


\section{Introduction}
\label{sec:introduction}

Our current understanding of massive star-forming regions remains poor, despite their importance in the structure and evolution of galactic systems, due to their strong feedback. From an observational point of view, a number of reasons make the study of massive star formation a challenging topic in astrophysics, as pointed out in the comprehensive review on the issue by \citet{Zinnecker07}. Regions of massive star formation are highly embedded in thick layers of dust during the crucial early stages of their existence. In addition, these early stages are short-lived, leaving little time for the study of their evolution. Another problem is the lack of spatial resolution in the observations. Most of extragalactic giant star forming regions are often contained within a single pixel. Finally, most massive stars form in close proximity to each other, and their mutual influence via gravitational interaction, powerful outflows, supernova events and strong winds contributes to the complexity of the problem. 

Observations of nearby star forming regions provide an excellent laboratory for the study of a particular aspect of this interaction, namely the triggering of new star formation events by gas compression resulting either from a ionization shock front created by the radiation field of a previous generation of stars, or by the supersonic shocks of a supernova event. In the classical theory by \citet{Elmegreen77}, an ionization front compresses an adjacent layer of molecular gas and heats it, producing a gravitational instability that will eventually result in a new generation of stars. In order to test this and other theories of triggered star formation, it is important to unequivocally determine and quantify the amount of currently ongoing star formation in Giant \hii\ Regions (G\hii Rs), and its relation to the ionizing radiation from previous generations of stars. 

After 30 Doradus, NGC~604 is the second most massive G\hii R in the Local Group. Located in the Triangle Galaxy (M33) at a distance of 0.84~Mpc \citep{Freedman91}, it harbors several associations of massive stars distributed across an area of about 200~pc on the side, dominated by a cluster containing $\sim 200$ OB stars \citep[][ and references therein]{Hunter96}. These associations have excavated a complex system of filaments and cavities of ionized material surrounded by photon-dominated regions (PDRs) and molecular gas \citep[see, for example][for a discussion on the spatial distribution of emission in the region]{Relano09}. Optical studies reveal an age of the region between 3 and 5 Myr \citep{Hunter96, Gonzalez_Delgado00} and a total stellar mass of $(3.8 \pm 0.6) \times 10^{5}\: \rm{M}_{\astrosun}$ \citep{Eldridge11}. Individual CO molecular clouds have been detected with sizes between 5 and 29~pc and with masses of between $0.8\times 10^{5}\: \rm{M}_{\astrosun}$ and $7.4\times 10^{5}\: \rm{M}_{\astrosun}$ \citep{Miura10}. Measured values of the average optical extinction in the region vary from $A_V = 0.3$ \citep{Relano09} to $A_V = 0.5$ \citep{Churchwell99}. \citet{Relano09} also derive a star formation of about $(5.7\pm 0.4)\times 10^5\: \rm{M}_{\astrosun}$, which can be interpreted in terms of the total mass of a coeval stellar population.  

Several attempts have been made to quantify the total amount of ongoing massive star formation in NGC~604. Using near-infrared (NIR) color-color diagrams, \citet{Barba09} identify several Massive Young Stellar Object (MYSO) candidates, with well-defined infrared excesses. They also note that these candidates coincide spatially with radio-peak structures, reinforcing their star-forming nature. More recently, \citet{Farina12} report the discovery of sources with near infrared excess within the infrared-bright ridges surrounding the ionized gas, and associate them with MYSOs. \citet{Relano09} argue that the reddening observed towards some prominent sources in the region can be explained by the existence of foreground molecular material, but they do not rule out the possibility of those sources being embedded sites of star formation. The use of infrared spectrometry combined with physical modeling provides a powerful tool to distinguish between foreground extinction and embedded star formation, by fitting the spectra with independent models that correspond to each situation, or to a combination of both. This complements photometric methods that use near-infrared color-color and color-magnitude diagrams to discriminate between the two cases, as the one described in \citet{Lada92}.

In this paper we perform a comprehensive analysis of the physical conditions in NGC~604 in the context of its evolutionary status, and investigate the presence of ongoing massive star formation in the region. We use infrared spectral and photometric data from the \emph{Spitzer} and \emph{Herschel} Space Telescopes, complemented with archival X-ray and optical data, as well as a set of analytical tools to interpret them. We report the discovery of individual infrared knots whose derived masses are consistent with them being stellar cluster in process of formation. This is also supported by Bayesian fitting of the Spectral Energy Distribution (SED) of the region, which points to the presence of a significant component of embedded objects in the region. We derive line emission and continuum maps of NGC~604 and use them to assess the variations in radiation field strength, ionization levels and extinction across the G\hii R, and find that our results are consistent with a sequential star formation history in the last $\sim 4\: $Myr. We also discuss some additional findings regarding the role of X-ray emission in the enhancement of both [\silii] atomic emission and 17~\mum\ PAH emission.

The paper is structured as follows. In \S \ref{sec:observations} we present the IRS data and describe the data reduction to obtain the maps and the spatially integrated spectrum of NGC~604. \S \ref{sec:analysis} describes the resulting maps and spectra, as well as the tools used to extract physical information from the observations. In \S \ref{sec:discussion} we discuss the results of our analysis in terms of the current evolutionary stage of NGC~604 and the presence of ongoing star formation in the region. We conclude with a summary of our main results in \S \ref{sec:conclusion}.

\section{Data Reduction and Ancillary Datasets}
\label{sec:observations}

Most of our analysis will be based on \emph{Spitzer} Infrared Spectrograph (IRS) \citep{Houck04} data of a region encompassing the bulk of infrared emission from NGC~604. However, to provide more robust constraints on the physics of the region, we also use complementary photometry from the \emph{Spitzer} Infrared Array Camera (IRAC) \citep{Fazio04} and from the \emph{Herschel Space Observatory} Photodetector Array Camera and Spectrometer (PACS) \citep{Poglitsch10}. Additionally, we use archival images from \emph{Hubble Space Telescope Wide Field and Planetary Camera 2} (WFPC2) and \emph{Chandra X-ray Observatory Advanced CCD Imaging Spectrometer} (ACIS).

\subsection{IRS data}

The \emph{Spitzer} IRS provides unprecedented spatial resolution and sensitivity as compared to any previous mid-IR spectroscopic observations of the NGC~604 region. The spectral resolving power ranges from $R\sim 60$ at short wavelengths to $R\sim 120$ at the long-wavelength edge. With a 512~s exposure, point source sensitivity limit between 0.1 and 10~mJy in the 5-38~\mum\ range, the IRS is about 100 times more sensitive than the \emph{Infrared Space Observatory} (ISO), while the spatial resolution is a factor of 10 larger. 

All the observations presented in this section were obtained in January 2006 using the IRS mapping and staring modes for spectroscopy, as part of the program \textit{Comparative Study of Galactic and Extragalactic \hii\ Regions} (P.~I.~J. Houck). The mapping mode consists of the acquisition of slit spectra using a grid of positions around a central target. Only the low resolution modules short-low (SL1, SL2) and long-low (LL1, LL2) of the IRS were used for the NGC~604 map presented in this paper. For the SL modules, 12 slit pointings were made with each of the two spectrometer orders covering an area on the sky of about $55^{\prime\prime}\times 40^{\prime\prime}$, which corresponds to a physical scale of about $225\: \rm{pc}\times 160~\rm{pc}$ at the distance of NGC~604 (see Fig.~\ref{fig:n604_bg}). The slice width for the SL modules is $3^{\prime\prime}.6$, corresponding to a pixel scale of $1^{\prime\prime}.85~\rm{px}^{-1}$ ($7.5~\rm{pc}~\rm{px}^{-1}$). For the LL modules, the slice width is $10^{\prime\prime}.5$, which corresponds to a pixel scale of $5^{\prime\prime}.08\: \rm{px}^{-1}$ (20.5$\: \rm{pc}\: \rm{px}^{-1}$) and 6 pointings were made with each order to cover an area of about $720\: \rm{pc}\times 205\: \rm{pc}$. 

In addition to the spectral map with the IRS low resolution modules (lores data hereafter), staring mode spectra were also obtained with the IRS for three specific locations within the region (hires data here after). In this ``point and shoot'' mode, the targets are placed in the center of one or several of the slits for a specified integration time. In staring mode, only the high resolution modules of the spectrometer were used, namely short-high (SH) and long-high (LH). The wavelength coverage is shorter (between 9.9~\mum\ and 38.0~\mum), but the resolving power is significantly higher ($R \approx 600$). The locations of the staring mode observations are within the area of the spectral map and correspond approximately to the positions of the peaks of 8~\mum\ emission, associated with PAH emission, as discussed later. The high resolution slits are wider than their low resolution counterparts ($4^{\prime\prime}.7$ and $11^{\prime\prime}.1$ for SH and LH, respectively); therefore their spectral apertures are also larger than the pixel scale of the lore spectral map. 

The area of the lores map and the locations of the staring mode hires spectra within the map are shown in the right panel of Fig.~\ref{fig:n604_rgb_map_first}.

\subsubsection{Extraction of the spectra}

For the staring mode data, the Spitzer Science Center Tool IRSCLEAN was used to remove cosmic rays. Then, SMART v.8.0 was used with the full aperture mode for extended sources, based on the size of the slits compared to the source NGC~604, which is large enough to be considered extended. The observation cycles were co-added and the sky background removed using an additional off-source background exposure taken as part of the campaign. We have applied a scaling factor to the fluxes extracted in the LH module to match the overlap region in the SH module. This is an aperture correction to account for the larger size of the LH slit with respect to the SH slit. The scaling factors for the LH spectra towards fields A, B, and C are 0.28, 0.41, and 0.41, respectively (see Fig.~\ref{fig:n604_rgb_map_first}).

\begin{figure}[h] \epsscale{1.1} \begin{center} \rotatebox{0}{\plotone{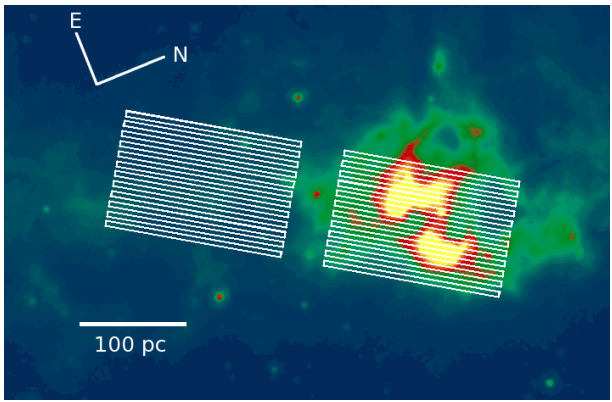}} \end{center} \caption{\label{fig:n604_bg} The target and background SL slit pointings superimposed on the IRAC 8$\: \mu$m image of NGC~604. Color coded are the IRAC fluxes from about $\sim 8$~MJy~$\rm{sr}^{-1}$ (dark blue) to $\sim 120$~MJy~$\rm{sr}^{-1}$ (yellow).  Emission from the spiral arm of M33 is visible towards the south-east of the \hii\ region, in the region where the background was taken. See text for a discussion on the flux levels.} \end{figure}

For the spectral map, we extracted the spatially integrated spectrum of the region for all four IRS orders using a extraction aperture corresponding to the SL coverage of the map (see Fig.~\ref{fig:n604_rgb_map_first}), which includes the bulk of the IR emission of NGC~604. Some of the extended filamentary structure of the region, which is one order of magnitude dimmer than the peaks of emission, is left outside this area. The low surface brightness of these filaments and the bubble-like geometry that we will assume when modeling the region implies that this will be unimportant for the purposes of the present analysis. The background subtraction was performed using the order of the SL and LL modules that was not centered at the source during the corresponding exposure. CUBISM performs the data cube build-up once the background and the correct slit pointings have been provided. The background levels are not exactly the same for the SL and LL modules. The reason is that the orientation of opposite orders of the slit is different for different modules. In the case of the low resolution modules, the background picks up some emission from M33's spiral arm (See Fig.~\ref{fig:n604_bg}). This background emission from the spiral arm is our best estimate for the sky levels in the SL modules. For reference, the measured IRAC 8~\mum\ flux levels on the spiral arm ($\sim 10\: $MJy~sr$^{-1}$) are only 20\% higher than the sky level outside the arm ($\sim 8\: $MJy~sr$^{-1}$), and correspond to about 7\% of the peak of 8~\mum\ emission in the region.

Spectra can be extracted from each spatial pixel of the resulting data cube. The spatially integrated spectrum over the entire aperture is obtained by summing up the individual spectra of all resolution elements. We use the same extraction area for all modules and orders using the tools provided by CUBISM to make sure that the final spectrum for each order corresponds to the same physical region. 

In addition to the spectrum of the integrated region, we have extracted the lores spectra of individual sources that we discuss in \S \ref{sec:individual_sources}. For this extraction we use an aperture of a single SL pixel (1$^{{\prime}{\prime}}$.85). The CUBISM software allows us to extract the spectra in the LL spectral module with this small aperture, by scaling down the fluxes of the larger LL pixels to the SL sizes. This is equivalent to applying a scaling factor to match the different orders of IRS spectra, and introduces additional flux uncertainties in the long wavelength modules, where the PSF is not well sampled. Nonetheless, we consider this scaling a good approximation of the actual fluxes at smaller scales, since the SL pixels are not sufficiently small to resolve the sources, and the LL pixels are not large enough to include more than one source.

By selecting a single pixel aperture, we have chosen to loose some spatial information on the individual sources, because a single pixel samples only half of the PSF FWHM, and in exchange we avoid off-source flux contamination. More important than a full sampling of the PSF for our analysis, are the variations of the PSF with wavelength, that might introduce artifacts in the measured spectral features. Using calibration data, \citet{Pereira_Santaella10} characterized the PSF variations with wavelength for the IRS reconstructed PSFs. Apart from an undulating behavior of the PSF size with wavelength, that they attribute to alignment issues and to the reconstruction algorithm used,  they compute variations of less that 10\% on the PSF FWHM for the SL module. Such variations are below the observational errors in our data, described in \S \ref{sec:pacs_photo} and \ref{sec:continuum_em}. The PSF of the LL module is affected by fringing and is difficult to characterize.

\subsection{IRAC photometry}
As part of the same Spitzer program, IRAC maps of the NGC~604 region were obtained at 3.6~\mum, 4.5~\mum, 5.8~\mum, and 8~\mum. The pixel scale for these maps is $1.2^{\prime\prime}~\rm{px}^{-1}$, and they cover an area of about $5^{\prime}\times 9 ^{\prime}$, several times larger than the area covered by the spectral map. For the purpose of this paper we have extracted the integrated flux of each map within a rectangular aperture area equal to the extraction area of the IRS spectral map. We perform this extraction using the FUNTOOLS package for the SAO \emph{ds9} software. The sky background is estimated from the map by measuring the flux in a box of the same size as the map, but shifted to the west, to an area where no source emission is observed. In Table \ref{tab:irac_phot} we list the measured fluxes. The listed uncertainties correspond to absolute flux calibration uncertainties ($\sim 3\%$), which are derived for point sources taking several systematic effects into account, such as array position dependence, pixel phase dependence, color correction, and aperture correction, as described in \citet{Reach05}. 

\begin{deluxetable}{cc}
\tablecolumns{2}
\tablecaption{Integrated IRAC photometry of NGC~604}

\tablehead{
  \colhead{Wavelength [\mum]} &
  \colhead{Flux [Jy]} 
}
 \startdata

3.6 & 0.067(0.002) \\
4.5 & 0.062(0.002) \\
5.8 & 0.322(0.010) \\
8.0 & 0.922(0.028) \\

 \enddata

\label{tab:irac_phot}
\end{deluxetable}

The IRAC maps provide a sharper view of the region as compared to the IRS spectral map at specific wavelengths, and will allow the identification of interesting sources.

\subsection{PACS photometry}
\label{sec:pacs_photo}
Imaging maps of the host galaxy M33 have been obtained with the Herschel PACS instrument using the green (100~\mum) and red (160~\mum) filters as part of the HERM33ES Herschel key project, and comprehensive description of the data reduction and obained maps can be found in \citet{Boquien11}. The maps were obtained with a slow scan speed of $20^{\prime\prime}~\rm{s}^{-1}$, and cover a total area of about 70$^{\prime} \times 70 ^{\prime}$. Here we use the integrated photometry of an area of the PACS maps equivalent to the size of the IRS spectral maps (right panel of Fig.~\ref{fig:n604_rgb_map_first}). The pixel sizes are 3$^{{\prime}{\prime}}$.2 for the green band and 6$^{{\prime}{\prime}}$.4 for the red band; about 2 and 4 times the IRS-SL pixel size. The obtained fluxes, integrated over the entire area of the map, are $F_{100\mum\ }=39.7\pm 4.0$~Jy and $F_{160\mum\ }=30.1\pm 3.0\: $Jy. The 10\% uncertainty comes from a combination of absolute calibration errors, uncertainties associated with differences in the PSF at 100~\mum\ and 160~\mum, and different pixel sizes in the two bands that lead to aperture uncertainties. The rms noise levels of the PACS maps are 2.6~mJy~px$^{-2}$ and 6.9~mJy~px$^{-2}$.

\subsection{HST-WFPC2 F555W data}
We use the optical images obtained at 0.55~\mum\ with the \emph{Hubble} WFPC2 using the F555W filter, described in \citet{Hunter96}. At angular resolutions of $0^{\prime\prime}.1$, this optical map reveals the location of the massive ionizing clusters that provide the radiative input for the NGC~604 system.

\subsection{Chandra X-ray Observatory-ACIS data}
We use archival ACIS X-ray data. The data were taken as part of the Chandra proposal ``The Giant Extragalactic Star-Forming Region NGC~604'' (Proposal ID 02600453, P.I. F.Damiani), and consist of a soft (0.5-1.2~keV) X-ray image of the entire nebula, with an exposures time of 90~ks. The pixel scale is $1^{\prime\prime}\: \rm{px}^{-1}$.

\section{Analysis}\label{sec:analysis}

We analyze the multi-wavelength observations described above using a set of analytical and statistical tools that are based on physical models of the region, and that we will describe shortly. The models compute the radiative transfer of the UV radiation as it traverses the ionized gas and molecular material around the \hii\ region. They also compute the dynamical evolution of its the expanding \hii\ region. We use these tools to derive physical properties of the region, such as dust temperatures, total stellar mass, hardness of the radiation field, and ionization state of the gas. In this section we present the obtained maps and spectra and describe the analytical tools that we use.


\subsection{Distribution of the emission}\label{sec:distribution}

\subsubsection{Overall distribution}
\begin{figure*}
  \centering
  \subfigure[]{
\includegraphics[scale=0.45,angle=0]{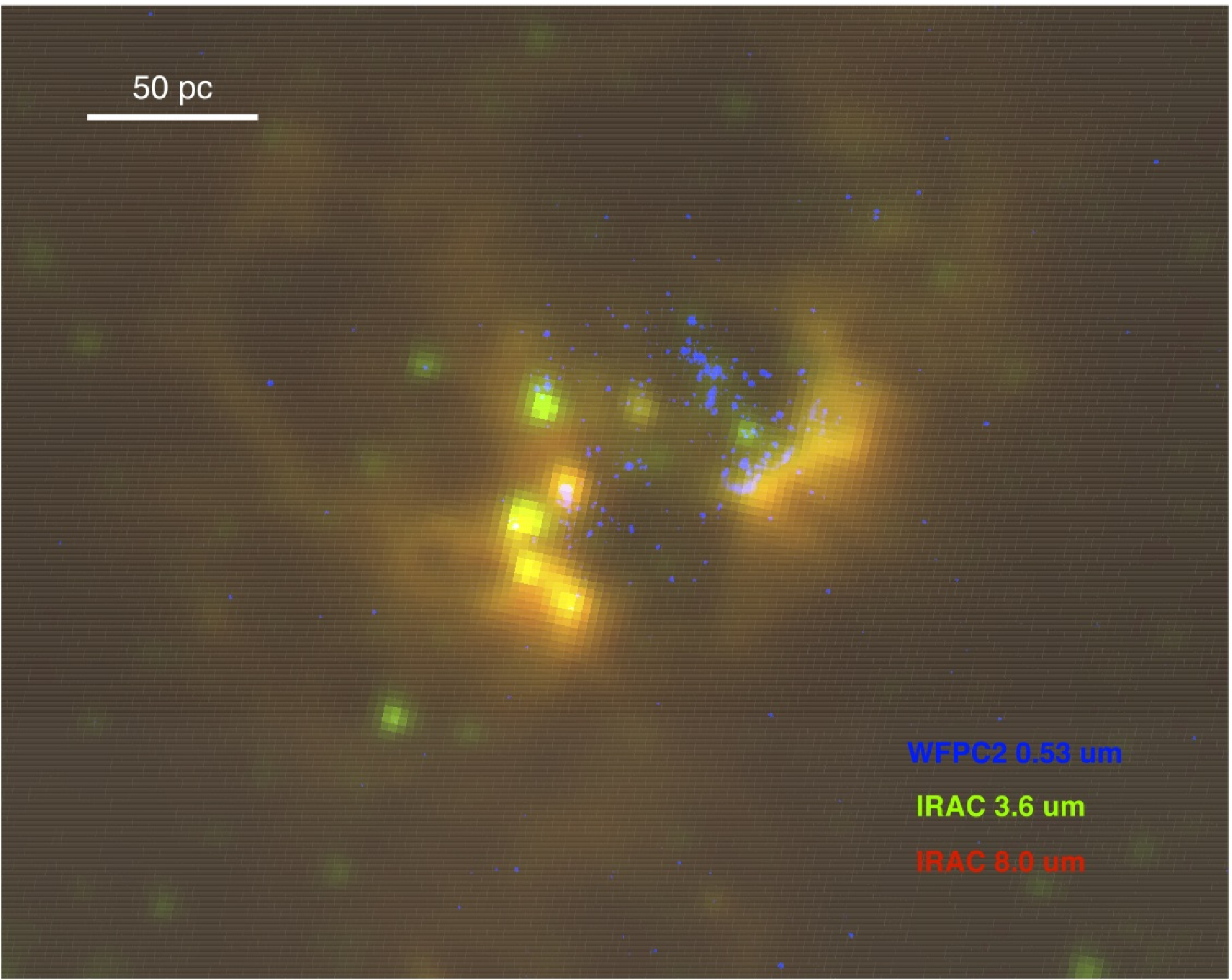}
\label{fig:n604_rgb_map_first_a}
}
  \subfigure[]{
\includegraphics[scale=0.45,angle=0]{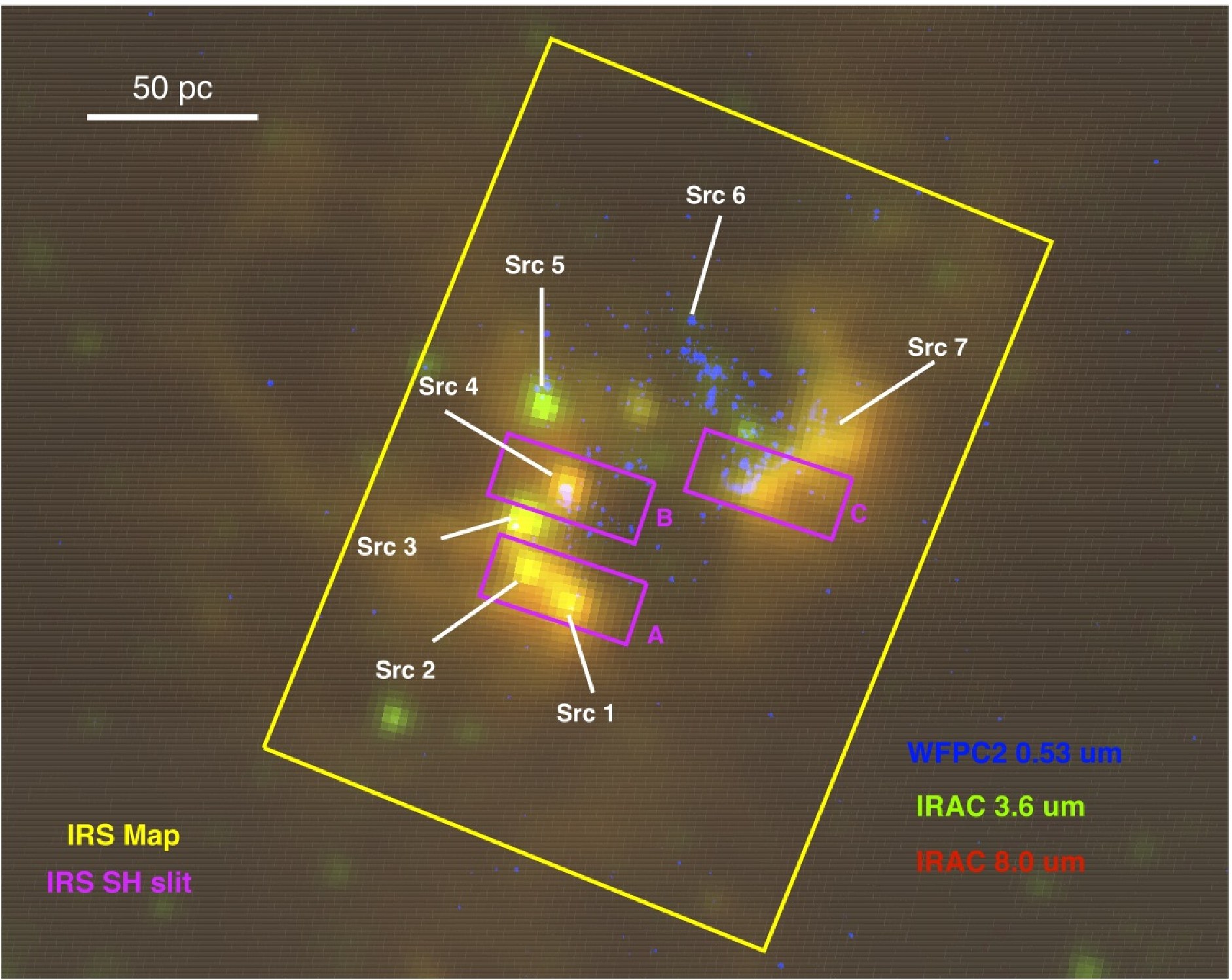}
\label{fig:n604_rgb_map_first_b}
}

  \caption{\emph{(a)} Three-color Hubble/Spitzer map of NGC~604. The IRAC 8~\mum\ band traces the PDR material, while the 3.6~\mum\ band traces both PAH and photospheric emission from young stars. The WFPC2 image shows the location of the hot massive stars. \emph{(b)} the same map with superimposed apertures of the IRS slits (A, B, and C) and labels for seven sources of interest based on their IRAC colors. We discuss these sources in the text and indicate their location in the map. The yellow rectangle corresponds to the extraction window for the IRS map and for the photometry, while the smaller magenta boxes show the field location of the IRS SH slit for the high resolution observations. The LH slits had the same locations as the SH slits. North is up, east is to the left.}

\label{fig:n604_rgb_map_first}
\end{figure*}

In Fig.~\ref{fig:n604_rgb_map_first_a} we show a three-color map of NGC~604 composed from the 8.-~\mum\ (red) and 3.6~\mum\ (green) IRAC channels together with the WFPC2 image at 0.55~\mum\ (blue) . The blue channel shows the photospheric emission from young massive stars, most of which belong to the dense clustered structure labeled as ``cluster A'' by \citet{Hunter96}. Other, more spread stellar associations are seen next to the bright lobes of infrared emission, most notably within fields B and C. The 8~\mum\ emission traces warm dust and the 7.7~\mum\ PAH feature, and hence the location of the PDRs, while the 3.6~\mum\ traces the 3.3~\mum\ PAH feature and photospheric emission from stellar populations older than 10~Myr, mostly giants \citep{Helou04, Cannon06}. The filamentary and shell-like structure of the region observed at optical wavelengths is reproduced in the mid-infrared. 

\citet{Relano09} have shown that the 8~\mum\ emission delineates the H$\alpha$ shells in the boundaries between cavities. Most of the infrared emission in our maps comes from two lobes oriented in a SE-NW direction and that constitute the bulk of the emission at the IRAC bands. These lobes coincide with the edges of two main cavities observed in the region \citep[cavities B1 and B2 in][]{Tullmann08}, with diameters of approximately 50~pc each. They also coincide with the position of bright radio knots identified by \citet{Churchwell99}. Within these cavities sits the majority of the luminous stars observed by HST (blue stellar sources in Fig.~\ref{fig:n604_rgb_map_first}). An exhaustive X-ray study of the region carried out by \citet{Tullmann08} showed that the distribution of high energy photons inside these cavities is consistent with them being shaped by the mass loss and radiative pressure of about 200 OB stars. The eastern portion of NGC~604, on the other hand, seems to be an older part of the system, with X-ray emission consistent with a more evolved population. Based on optical spectroscopy, also \citet{Tenorio_Tagle00} report a kinematical difference between the eastern and western parts of NGC~604 and claim that the eastern part is not affected by the ionized cluster.

The two lobes have similar surface brightness at 8~\mum\, but they show differences in their morphology. At the IRAC wavelengths, the SE lobe shows a series of sub-condensations, some of which are notably brighter at 3.6~\mum\, while the NW lobe shows a more uniform distribution of emission. CO maps of the region reveal several molecular clouds in the region, with a CO-bright cloud peaking near the SE infrared lobe and extending southwards, and a dimmer and smaller cloud peaking near the NW lobe \citep{Wilson92}. A number of MYSO candidates have been identified along these two lobes as sources with NIR excess \citep[see, for example][]{Farina12}. By combining observational and SED modelling tools, in this chapter we will provide additional evidence that supports the star formation scenario, and quantifies its contribution to the total stellar mass in the region.

\subsubsection{Individual Sources.}\label{sec:individual_sources}
\label{sec:indiv_sources}

Fig.~\ref{fig:n604_rgb_map_first_b} shows a three color map of NGC~604 with the main extraction areas indicated (see Table \ref{tab:sources_coord}). The extraction area for the spectral map includes the bulk of the emission in the IRAC bands and the stellar cluster. We have selected 7 individual sources of interest in the spectral map, including the well defined sub-condensations that are visible in the IRAC images. These sources are indicated in Fig.~\ref{fig:n604_rgb_map_first_b}, with their positions listed in Table \ref{tab:sources_coord}. Fields A and B of the hires observations include some of the selected sources of the lores map. Specifically, field A includes sources 1 and 2, while field B includes source 4. Field C does not include any of the selected lores targets, but belongs to the NW infrared lobe, and is close to source 7. Besides, fields A, B, and C correspond to radio fields B, A, and C, respectively, of \citet{Churchwell99}, while source 6 is slightly shifted from the cluster A of \citet{Hunter96}.

We have fitted Gaussian profiles to the flux distribution of the observed sub-condensations to investigate their spatial extension due to the PAH emitting regions. Our fits reveal that they have projected diameters of about $3^{\prime\prime}.6$ at 8~\mum, as measured from the FWHM of the Gaussian fits. The diffraction-limited resolution of the IRAC camera at this wavelength is $1^{\prime\prime}.71$, and hence we conclude that the sub-condensations are spatially resolved at 8~\mum, having a diameter of at least two IRAC resolution elements. At the distance of NGC~604, their projected sizes correspond to physical diameters of about 15~pc; hence, comparable to the size of a typical galactic giant molecular cloud \citep{Fukui10}.

\begin{deluxetable}{ccc}
\tablecolumns{3}
\tablecaption{IRS targets in NGC~604\\Lores sources and hires fields}

\tablehead{
  \colhead{Source} &
  \colhead{RA} &
  \colhead{Dec} 
}
\startdata

Src 1 & 1h34m33.43s & $+30^{\circ}46^{\prime}48.50^{\prime\prime}$ \\
Src 2 & 1h34m33.67s & $+30^{\circ}46^{\prime}51.01^{\prime\prime}$ \\
Src 3 & 1h34m33.70s & $+30^{\circ}46^{\prime}54.86^{\prime\prime}$ \\
Src 4 & 1h34m33.43s & $+30^{\circ}46^{\prime}57.11^{\prime\prime}$ \\
Src 5 & 1h34m33.56s & $+30^{\circ}47^{\prime}03.00^{\prime\prime}$ \\
Src 6 & 1h34m32.73s & $+30^{\circ}47^{\prime}09.32^{\prime\prime}$ \\
Src 7 & 1h34m31.99s & $+30^{\circ}46^{\prime}59.95^{\prime\prime}$ \\
A & 1h34m33.6s & $+30^{\circ}46^{\prime}51^{\prime\prime}$ \\
B & 1h34m33.6s & $+30^{\circ}46^{\prime}58^{\prime\prime}$ \\
C & 1h34m32.4s & $+30^{\circ}46^{\prime}59^{\prime\prime}$ \\
\enddata

\label{tab:sources_coord}
\end{deluxetable}

\subsection{Infrared Spectra}
\label{sec:spectra}

\subsubsection{Extracted Spectra}
\begin{figure}
  \centering
  
  \includegraphics[scale=0.63,angle=0]{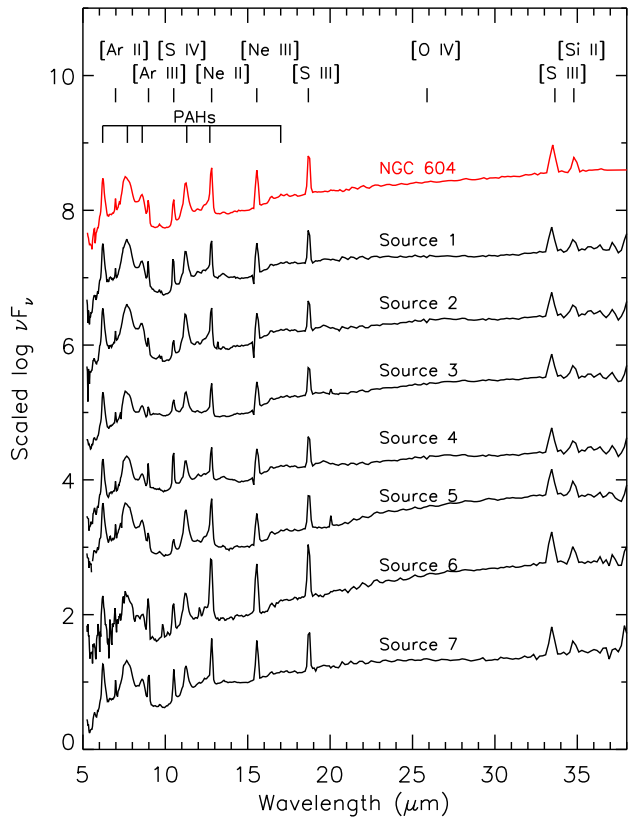}
  
  \caption{IRS spectra of the integrated NGC~604 region and the individual sources listed in Table \ref{tab:sources_coord}. The position of some prominent mid-infrared fine structure lines are indicated, as well as the location of the PAH bands. The small feature seen at $\sim$20~\mum\ in sources 3 and 5 is an artifact from the data reduction.}
    
  \label{fig:all_spectra}
\end{figure}

In Fig.~\ref{fig:all_spectra} we show the lores IRS spectra of the integrated region and the selected sources of Table \ref{tab:sources_coord}. The vertical axis is in units of flux density ($\nu F\nu$), and the spectra have been scaled by progressive factors of 10 to show the direct comparison. The integrated spectrum of NGC~604 has prominent PAH emission and some of the nebular lines detected are from species such as [\ariii], [\neiii], [\neii], [\siv], [\siii], and [\silii]. The thermal continuum increases monotonically in the IRS range.

\begin{figure}
  \centering
  
  \includegraphics[scale=0.75,angle=0]{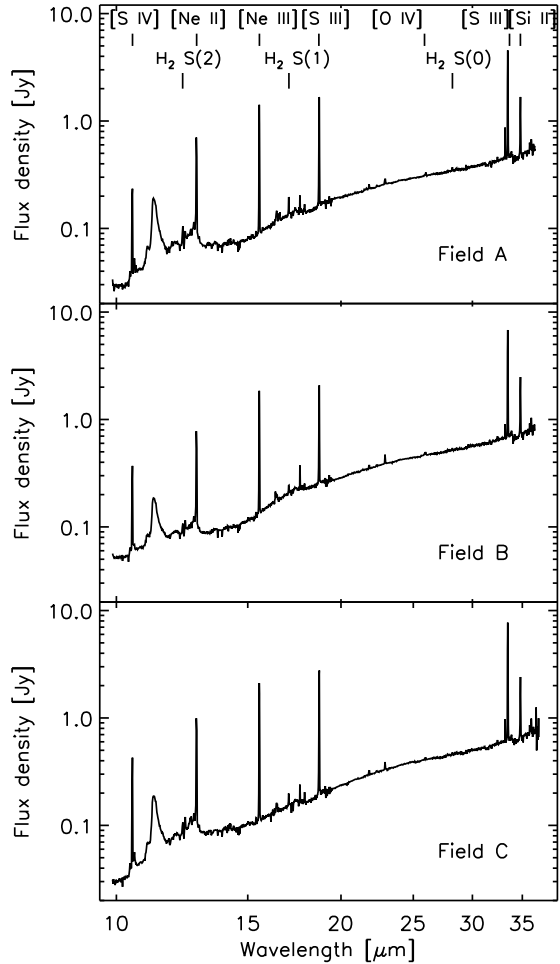}
  
  \caption{High resolution spectra of the three fields labelled A, B, and C in Fig.~\ref{fig:n604_rgb_map_first}. The prominent nebular and molecular hydrogen lines are indicated.}
  
  \label{fig:hires_spec}
\end{figure}

Fig.~\ref{fig:hires_spec} shows the high resolution spectra of the staring mode targets. The wavelength coverage of the hires modules is smaller than the lores case, from $\sim 10~\mum$ to $\sim 37~\mum$, but the higher spectral resolving power allows the resolution of lines not seen in the lores spectra, such as the lines resulting from pure rotational transitions of molecular hydrogen, H$_2\: $S(2) at 12.3~\mum\ and H$_2\: $S(1) at 17.0~\mum. These lines are usually hard to detect on top of a strong continuum, due to the fact that they arise from quadrupolar rotational transitions, which are intrinsically weak. 

\subsubsection{Continuum emission}
\label{sec:continuum_em}
In order to characterize the spectral slope of the thermal continuum, we measure the flux densities at 15~\mum\ and 30~\mum\ using a range of wavelengths around the corresponding central wavelength containing about 20 resolution elements (14.75~\mum\ -15.25~\mum\ for the 15~\mum\ measurement and 29.5~\mum\ -30.5~\mum\ for the 30~\mum\ measurement). We calculate the ratio $\rm{F}_{15\rm{\mum}}/\rm{F}_{30\rm{\mum}}$ for each source as well as the integrated spectrum. The spectral slopes give an indication of the dust temperature. Higher values of $\rm{F}_{15\mum}/\rm{F}_{30\rm{\mum}}$ are associated with a hotter component of the dust, whereas lower values of this ratio indicate colder dust temperatures. We list the measured fluxes and slopes in Table \ref{tab:cont_slope}. We have used a nominal 10\% uncertainty in the IRS fluxes, which accounts for the absolute and relative calibration of the instrument, as well as systematic errors due to specific observing conditions.

\begin{deluxetable}{cccc}
\tablecolumns{4}
\tablecaption{SED continuum slope for all seven sources and the integrated map.}

\tablehead{
  \colhead{Source} &
  \colhead{F$_{15\mum}$} &
  \colhead{F$_{30\mum}$} &
  \colhead{F$_{15\mum}$/F$_{30\mum}$} \\
  \colhead{} &
  \colhead{[MJy~sr$^{-1}$]} &
  \colhead{[MJy~sr$^{-1}$]} &
  \colhead{} 
}
 \startdata

Src 1 & 102.8(10.3) & 432.1(43.2) & 0.238(0.034) \\
Src 2 & 65.4(6.5) & 310.4(31.0) &  0.211(0.030) \\
Src 3 & 95.8(9.6) & 582.8(58.3) &  0.164(0.023) \\
Src 4 & 95.9(9.6) & 447.5(44.8) &  0.214(0.030) \\
Src 5 & 25.8(2.6) & 258.5(25.9) &  0.100(0.014) \\
Src 6 & 10.1(1.0) & 99.2(9.9) &  0.101(0.014) \\
Src 7 & 113.5(11.4) & 493.9(49.4) & 0.230(0.033) \\
NGC~604 & 16.8(1.7) & 100.9(10.1) &  0.166(0.023) \\

\enddata

\label{tab:cont_slope}

\end{deluxetable}

\subsubsection{PAH emission}

We have measured the strengths of the individual PAH features using the PAHFIT tool \citep{Smith07}, which decomposes the IRS spectrum in individual contributions from dust thermal continuum, PAH features, and fine-structure lines, as well as rotational lines of molecular hydrogen. In Table \ref{tab:pah_features} we list the measured continuum-subtracted strengths of the different PAH components, and their uncertainties as derived with PAHFIT. The 7.7~\mum\ strength was obtained by adding the PAHFIT 7.4~\mum, 7.6~\mum, and 7.8~\mum\ features. Similar combinations were performed for the 11.3~\mum\ feature (using the 11.2~\mum\ and 11.3~\mum\ strengths) and for the 17~\mum\ feature (using the 16.4~\mum, 17.0~\mum, 17.4~\mum, and 17.9~\mum\ strengths).

\begin{deluxetable*}{cccccc}
\tablecolumns{6}
\tablecaption{PAH features strengths}

\tablehead{
  \colhead{Source} &
  \colhead{PAH$_{6.2\mum}$} &
  \colhead{PAH$_{7.7\mum}$} &
  \colhead{PAH$_{8.6\mum}$} &
  \colhead{PAH$_{11.3\mum}$} &
  \colhead{PAH$_{17\mum}$}  \\
  \colhead{ } &
  \colhead{$10^{-6}\times$[W~m$^{-2}$~sr$^{-1}$]} &
  \colhead{$10^{-6}\times$[W~m$^{-2}$~sr$^{-1}$]} &
  \colhead{$10^{-6}\times$[W~m$^{-2}$~sr$^{-1}$]} &
  \colhead{$10^{-6}\times$[W~m$^{-2}$~sr$^{-1}$]} &
  \colhead{$10^{-6}\times$[W~m$^{-2}$~sr$^{-1}$]}
}

 \startdata

Src 1 &  2.56(0.10) & 7.96(0.54) & 1.28(0.12) & 1.33(0.16) & 0.85(0.26)\\
Src 2 &  1.76(0.07) & 5.49(0.38) & 0.87(0.09) & 1.07(0.11) & 0.73(0.17)\\
Src 3 &  1.24(0.07) & 2.82(0.33) & 0.51(0.09) & 0.83(0.14) & 1.15(0.24)\\
Src 4 &  1.51(0.07) & 4.25(0.35) & 0.64(0.09) & 0.76(0.13) & 0.88(0.23)\\
Src 5 &  0.93(0.03) & 2.42(0.19) & 0.50(0.04) & 0.52(0.05) & 0.24(0.07)\\
Src 6 &  0.14(0.01) & 0.48(0.03) & 0.06(0.01) & 0.11(0.01) & 0.06(0.03)\\
Src 7 &  1.64(0.08) & 5.14(0.41) & 0.91(0.10) & 1.04(0.14) & 0.26(0.07)\\
NGC~604 &  0.40(0.02) & 1.16(0.09) & 0.19(0.02) & 0.25(0.03) & 0.11(0.04)\\
    
 \enddata

\label{tab:pah_features}
\end{deluxetable*}

In general, the ratios between PAH feature strengths  have similar values for all sources, with source-to-source variations of between 6\% and 14\%. An interesting exception is the ratio of the PAH$_{17\mum }$ feature to the sum of all the other PAH features, $\Sigma \rm{PAH}$ ($\Sigma PAH = \rm{PAH}_{6.2\mum }+\rm{PAH}_{7.7\mum }+\rm{PAH}_{8.6\mum }+\rm{PAH}_{11.3\mum }$). This ratio shows source-to-source variations of about 60\% and peaks strongly towards source 3, which is also a local peak of soft X-ray emission. In Table \ref{tab:pah_ratios} we list the PAH$_{17\mum }/\Sigma \rm{PAH}$ ratio and the Chandra-ACIS soft X-ray fluxes for our sources. For the ACIS fluxes we assume a nominal 20\% uncertainty, based on Chandra catalogues of point sources such as \citet{Servillat08}.

\begin{deluxetable}{ccc}
\tablecolumns{3}
\tablecaption{Relation between the ratio of PAH 17~\mum\ to all the other PAH features and soft X-ray emission.}

\tablehead{
  \colhead{Source} &
  \colhead{PAH$_{17\mum}$/($\Sigma \rm{PAH}$)\tablenotemark{a}} &
  \colhead{$F_{X_{\rm{soft}}}$ [$10^{-9}\: $erg$\: \rm{cm}^{-2}\: \rm{s}^{-1}$]\tablenotemark{b}}
}

 \startdata

Src 1 &  0.065(0.020) & 0.84(0.17)  \\
Src 2 &  0.079(0.019) & 1.37(0.27)  \\
Src 3 &  0.213(0.047) & 3.70(0.74)  \\
Src 4 &  0.123(0.033) & 2.50(0.50)  \\
Src 5 &  0.055(0.016) & 1.64(0.33)  \\
Src 6 &  0.076(0.038) & 1.65(0.33)  \\
Src 7 &  0.030(0.002) & 0.91(0.18)  \\
NGC~604 &  0.055(0.020) & -  \\
    
 \enddata
\tablenotetext{a}{Fluxes are from PAHFIT. $\Sigma \rm{PAH}$ is obtained as the sum of the PAH features at 6.2~\mum, 7.7~\mum, 8.5~\mum, and 11.3~\mum.}
\tablenotetext{b}{Measured from \emph{Chandra}-ACIS data.}

\label{tab:pah_ratios}
\end{deluxetable}

\subsubsection{Nebular line emission}

Fine-structure emission lines are an important diagnostic of the physical conditions in star-forming regions. Their strengths and ratios constrain physical parameters such as the hardness of the radiation field, gas density, or the presence of shocked gas. In the mid-IR, several forbidden lines are observable from species such as [\neii], [\neiii], [\siii], and [\siv]. In Table \ref{tab:nebular} we list the continuum-subtracted nebular line strengths in NGC~604, as measured with the PAHFIT tool, as well as their uncertainties from the fit, also returned by PAHFIT.

We have also measured the continuum-subtracted line strengths in the hires data for targets A, B, and C in the right panel of Fig.~\ref{fig:n604_rgb_map_first}. We have fitted Gaussian profiles to the lines, this time using the built-in tool for that purpose included in the SMART software. We show the resulting line strengths in Table \ref{tab:hires_lines}.

\begin{deluxetable*}{ccccccccc}
\tablecolumns{9}
\tablecaption{Lores fine structure lines\tablenotemark{a}}

\tablehead{
  \colhead{Source} &
  \colhead{[\arii]$6.9\mum$} &
  \colhead{[\ariii]$8.9\mum$} &
  \colhead{[\siv]$10.5\mum$} &
  \colhead{[\neii]$12.8\mum$} &
  \colhead{[\neiii]$15.5\mum$} &
  \colhead{[\siii]$18.7\mum$} &
  \colhead{[\siii]$33.7\mum$} &
  \colhead{[\silii]$34.8\mum$} 
}

 \startdata

Src 1 &  0.58(0.19) & 2.05(0.33) & 3.07(0.42) & 4.61(0.64) & 4.83(0.67) & 6.41(0.91) & 6.64(1.13) & 2.54(0.76)\\
Src 2 &  0.60(0.17) & 0.58(0.16) & 0.84(0.14) & 2.89(0.40) & 2.81(0.40) & 3.53(0.54) & 4.29(0.76) & 1.74(0.53)\\
Src 3 &  0.59(0.14) & 0.90(0.34) & 1.54(0.28) & 3.18(0.39) & 3.84(0.47) & 5.66(0.83) & 7.90(1.40) & 3.29(0.96)\\
Src 4 &  0.48(0.16) & 2.05(0.31) & 4.02(0.53) & 3.39(0.52) & 4.08(0.48) & 5.37(0.79) & 6.52(1.13) & 2.57(0.76)\\
Src 5 &  0.51(0.08) & 0.47(0.09) & 0.32(0.06) & 1.85(0.19) & 1.15(0.13) & 2.21(0.23) & 4.37(0.72) & 1.90(0.47)\\
Src 6 &  0.15(0.02) & 0.35(0.04) & 0.25(0.03) & 1.28(0.11) & 1.02(0.11) & 1.65(0.01) & 2.34(0.01) & 0.90(0.01)\\
Src 7 &  0.93(0.21) & 2.59(0.35) & 2.68(0.45) & 6.80(0.86) & 7.02(0.94) & 9.10(1.30) & 8.90(1.11) & 3.15(0.74)\\
NGC~604 &  0.17(0.04) & 0.29(0.06) & 0.34(0.04) & 1.04(0.11) & 1.04(0.10) & 1.57(0.01) & 1.90(0.01) & 0.77(0.01)\\
    
 \enddata

\label{tab:nebular}
\tablenotetext{a}{Units are $10^{-7}\: $Wm$^{-2}$sr$^{-1}$}
\end{deluxetable*}

\begin{deluxetable*}{ccccccc}
\tablecolumns{9}
\tablecaption{Fine structure line fluxes as extracted from the hires data.}

\tablehead{
  \colhead{Source} &
  \colhead{[\siv]$10.5\mum$} &
  \colhead{[\neii]$12.8\mum$} &
  \colhead{[\neiii]$15.5\mum$} &
  \colhead{[\siii]$18.7\mum$} &
  \colhead{[\siii]$33.7\mum$} &
  \colhead{[\silii]$34.8\mum$} \\
  \colhead{ } &
  \colhead{[$10^{-20}\times$W~cm$^{-2}$]} &
  \colhead{[$10^{-20}\times$W~cm$^{-2}$]} &
  \colhead{[$10^{-20}\times$W~cm$^{-2}$]} &
  \colhead{[$10^{-20}\times$W~cm$^{-2}$]} &
  \colhead{[$10^{-20}\times$W~cm$^{-2}$]} &
  \colhead{[$10^{-20}\times$W~cm$^{-2}$]} 
}

 \startdata

Field A &  1.63(0.36) & 3.71(0.14) & 4.24(0.08) & 4.54(0.12) & 5.54(0.07) & 1.85(0.09) \\
Field B &   2.24(0.22) & 4.24(0.24) & 5.52(0.12) & 5.49(0.12) & 9.03(0.09) & 2.76(0.09) \\
Field C &  2.60(0.46) & 5.77(0.24) & 6.63(0.02) & 7.29(0.18) & 10.08(0.09) & 2.71(0.11) \\
    
 \enddata

\label{tab:hires_lines}
\end{deluxetable*}

\begin{figure*}
  \centering
  \subfigure[]{

\includegraphics[scale=0.45,angle=0]{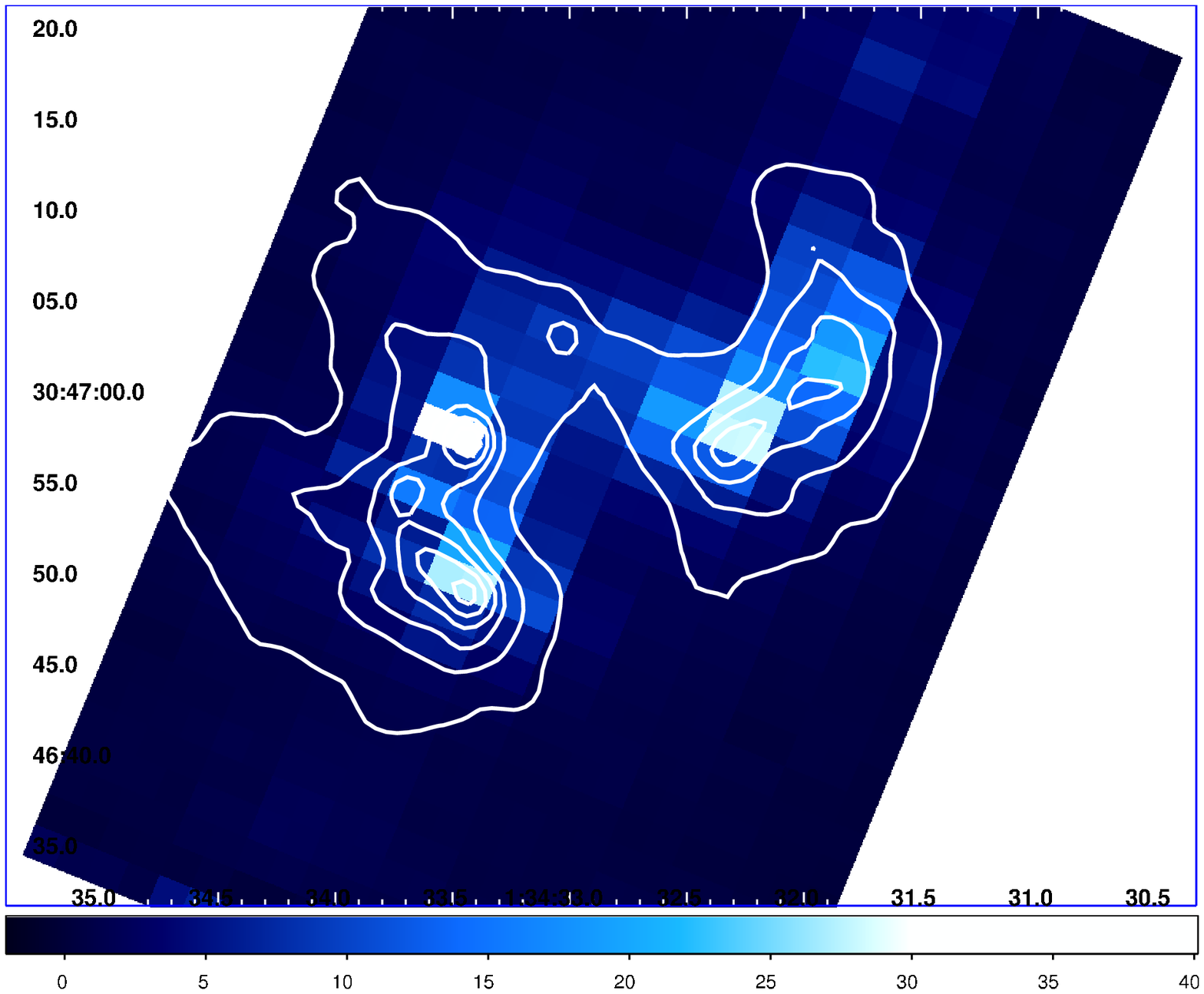}

}
  \subfigure[]{
\includegraphics[scale=0.45,angle=0]{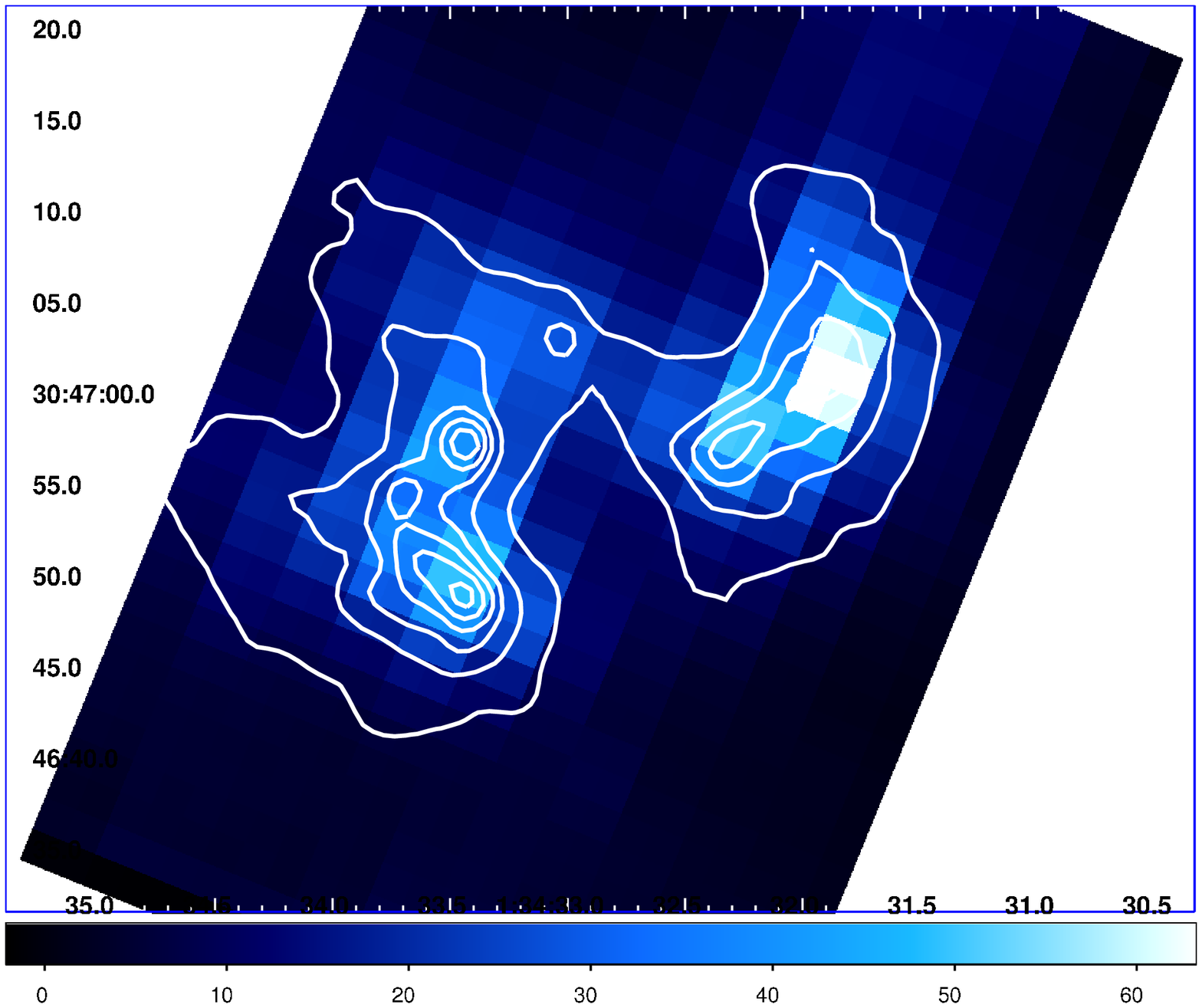}

}

  \caption{Continuum-subtracted line emission in NGC~604. Color coded are the IRS fluxes in MJy~sr$^{-1}$. \emph{(a) }[\siv]10.5\mum\ emission has well defined peaks near sources 1, 4, and field C. \emph{(b) }[\neii] peaks near source 7. The white contours are IRAC 8~\mum\ emission, and the RMS noise values are 0.45~MJy~sr$^{-1}$ and 4.5~MJy~sr$^{-1}$, respectively.}

\label{fig:line_maps}
\end{figure*}

Fig.~\ref{fig:line_maps} shows continuum-subtracted maps of the [\siv]10.5\mum\ and [\neii]12.8\mum\ fine-structure lines. Within the maps spatial resolution there is spatial coincidence between the emission line peaks and the infrared sources we have identified. Sources of particularly bright [\siv] line emission are sources 1 and 4, along with Field C. Source 7, on the other hand, is prominent in [\neii] emission. The localized nature of the emission allows investigations of the physical conditions in the individual sources.

\subsection{Electron density}
\label{sec:e_density}

The ratio of two lines of the same ionization state of a single species, emitted from levels with similar excitation energies can be used as a tracer of the electron density $n_e$ \citep{Rubin94}. We use the [\siii]18.7\mum/[\siii]33.6\mum\ to investigate the electron densities in the NGC~604 region, based on the lores line ratios. We use the lores data rather than the hires data because the two [\siii] lines are measured at similar spatial resolutions in the lores data, while for the hires data they are measured in two different apertures ([\siii]18.7 falls on the SH module, while [\siii]33.6 falls in the LH module). 

The [\siii]18.7\mum/[\siii]33.6\mum\ ratios calculated from Table \ref{tab:nebular} are very uniform in the region, varying from 0.51 to 1.02, with the highest value near source 7, in the NW infrared lobe. These values are very close to the low density limit discussed in \citet{Dudik07} (their Fig.~9), and hence our data provide only approximate upper limits for the electron densities. Using the \citet{Dudik07} diagram corresponding to a gas temperature of $10^4\: $K, we derive upper limits for the electron densities between $\log\: n_e=1.5\: \rm{cm}^{-3}$ and $\log\: n_e=2.5\: \rm{cm}^{-3}$ in the region. These values are in good agreement with the electron density derived by \citet{Maiz_Apellaniz04} using the optical line ratio [\sii]$\lambda 6717/\lambda 6731$.

\subsection{Ionization conditions}\label{sec:hardness}

\subsubsection{Hardness of the radiation field}

\begin{figure}
  \centering
  
  \includegraphics[scale=0.44,angle=0]{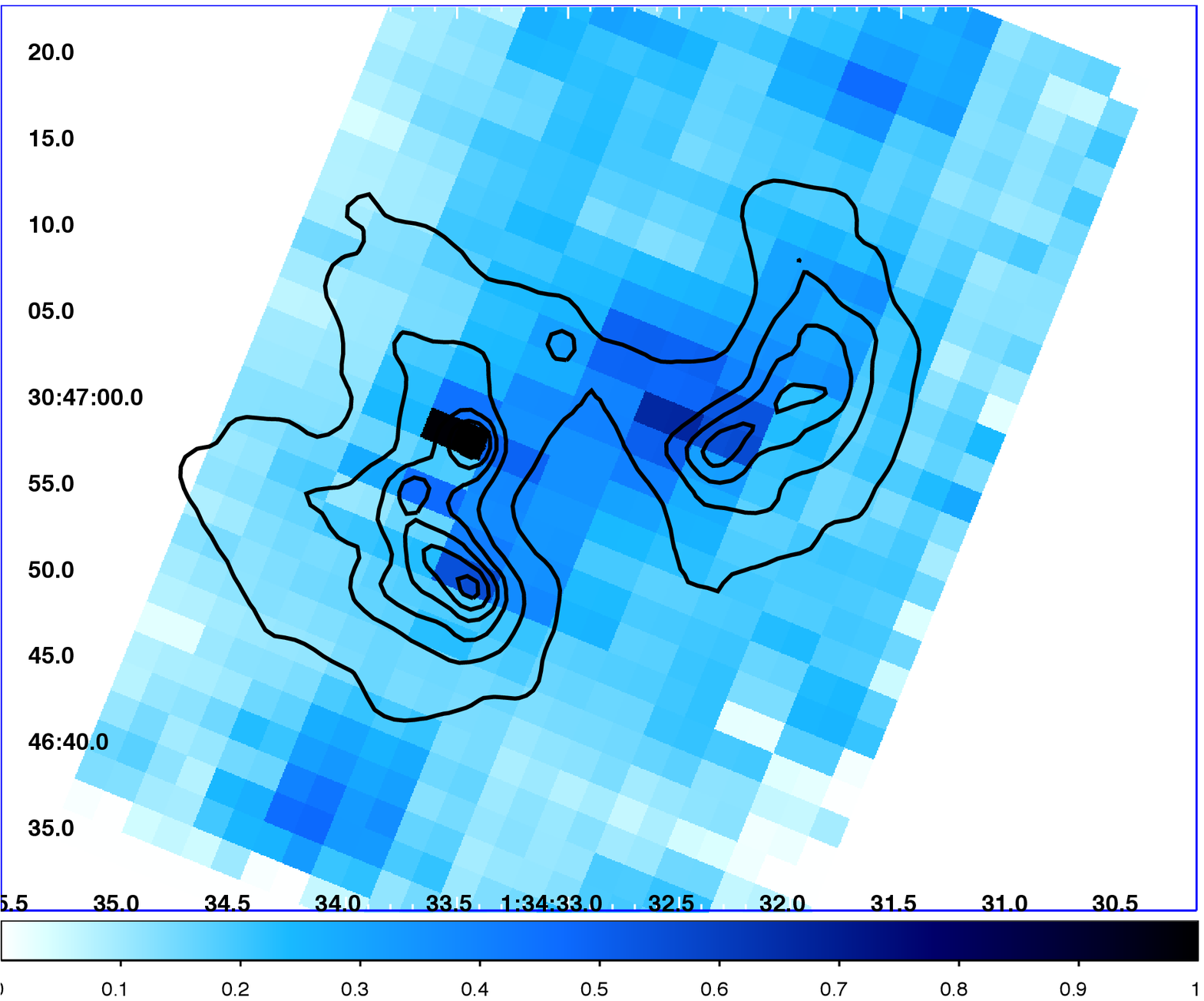}
  \label{fig:rgb_2}
   
  \caption{The [\siv]10.5\mum/[\neii]12.8\mum\ ratio in NGC~604 as derived from the spectral map. Color coded is the value of the ratio. The RMS noise equals 0.12. For reference, the IRAC 8~\mum\ contours are superimposed as the black solid line. The measured ratios are used to estimate the hardness of the radiation field towards specific locations (see text).}
  
  \label{fig:siv_over_nii}
\end{figure}

We use the lores [\siv]10.5\mum/[\neii]12.8\mum\ line ratio map to visually trace the spatial variations in the hardness of the radiation field. The use of these two lines has an observational motivation. The two lines are within the SL wavelength range and hence have the same spatial scale and pixel size in the map. This is not the case for the [\neiii]/[\neii] ratio, which was a more natural choice. The ionization potential required to produce [\neii] is only 21.6~eV, while it takes 34.8 eV to ionize [\siii] to obtain [\siv]. Hence, the ratio between these two ionic species traces the hardness of the radiation field, which can be interpreted in terms of stellar ages, with [\neii] tracing star formation activity during the last 10~Myr and [\siv] tracing massive stars born in the last 4-6 Myr. In Fig.~\ref{fig:siv_over_nii} we show the corresponding line ratio map obtained from the spectral cube. We further discuss this map in \S \ref{sec:sources741}.

\subsubsection{Ionization state of the gas}
The mid-IR line ratios do not only depend on the hardness of the radiation field. They also depend on the ionization state of the gas, that can be parametrized using the ionization parameter $Q$ (the ratio of the ionizing photon density to gas density). In \citet{Martinez_Galarza11} (MG11 hereafter) we have used the hires line ratios as derived from the measurements done in \citet{Lebouteiller08}, and combined them with the radiative transfer models in \citet{Levesque10} using the interactive ITERA software \citep{Groves10}, to break the degeneracy between age and ionization parameter in the particular case of the 30 Doradus region. We found in that study that the [\neiii]15.5\mum/[\neii]12.8\mum\ and the [\siv]10.5\mum/[\siii]18.7\mum\ ratios in 30 Doradus are compatible with starburst ages $\leq 3.0\: \rm{Myr}$ and ionization parameters $8.0\: \rm{cm}\: \rm{s}^{-1} \leq \log Q \leq 8.6\: \rm{cm}\: \rm{s}^{-1}$. 

\begin{figure}
  \centering
  
\includegraphics[scale=0.5,angle=0]{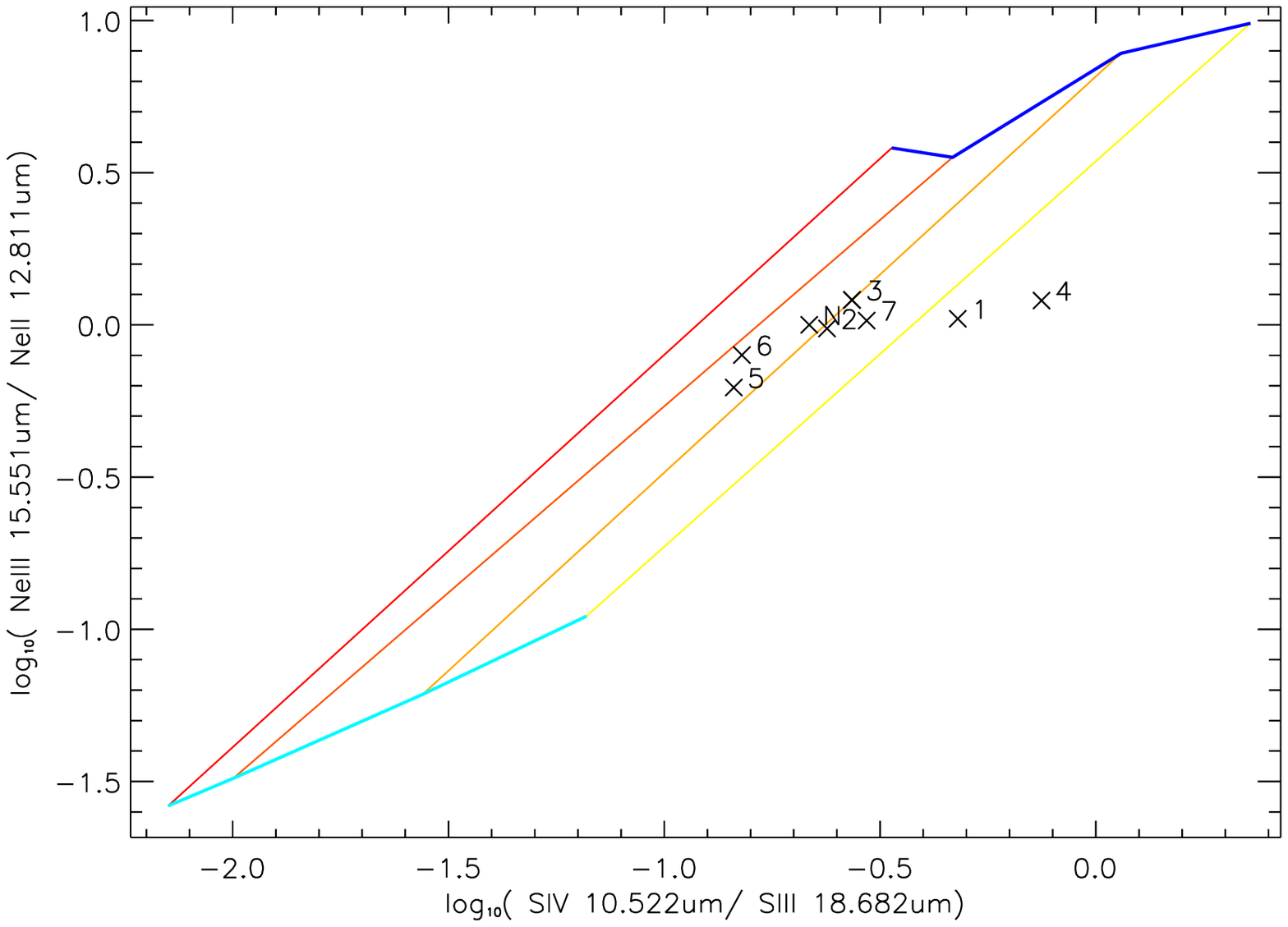}

  \caption{Comparison of the line ratios in NGC~604 with the \citet{Levesque10} models. The [\neiii]15.5\mum/[\neii]12.8\mum\ ratio is plotted versus the [\siv]10.5\mum/[\siii]18.7\mum\ ratio. Data points from Table \ref{tab:nebular} are overplotted to the grid of models, that cover a range of ages from 4~Myr (blue) to 4.5~Myr (cyan) and a range of ionization parameters from $8\times 10^7\: \rm{cm}\: \rm{s}^{-1}$ (red) to $4\times 10^8\: \rm{cm}\: \rm{s}^{-1}$ (yellow). The line ratio for the integrated map is indicated by ``N'', to the upper right of the symbol. Based on the propagation of the flux uncertainties listed in Table \ref{tab:nebular}, typical errors in the shown line ratios are 0.1 dex.}

\label{fig:itera}
\end{figure}

We perform a similar analysis for NGC~604 using the lores lines, which are extracted from the same spatial area regardless of the IRS module in which they fall. In Fig.~\ref{fig:itera} we plot line ratios computed from Table \ref{tab:nebular} as compared to a set of \citet{Levesque10} models of an instantaneous starburst with sub-solar metallicity ($Z=0.4Z_{\astrosun}$) and low electron density ($n_e=10\: \rm{cm}^{-3}$), in accordance with the low density regime inferred from the line ratios. The overall line ratios are indicative of ages between 4 and 4.5~Myr for the individual sources. Although both the observational uncertainties in the line ratios and the spatial confusion between the unresolved sources in the IRS map affect the measured line ratios, Fig.~\ref{fig:itera} indicates a similar age for all the infrared bright sources. However, sources 1 and 4 show larger ionization parameters, sources 2, 3, and 7 show moderated ionization parameters, and sources 5 and 6 have the lowest ionization parameters. While this is not necessarily indicative of a stronger radiation field near sources 1 and 4, it points to a large number of ionizing photons near these sources.

Using the hires line fluxes we also measure the strength of the radiation field in fields A, B, and C with the parametrization of \citet{Beirao06}, which uses the [\neiii]/[\neii] to estimate the strength as the product of the field intensity and the field hardness:

\begin{equation}
\label{field_strength}
\left(F_{[\rm{Ne\,{\sc II}}]12.8\mum}+F_{[\rm{Ne\,{\sc III}}]15.6\mum}\right) \times \frac{F_{[\rm{Ne\,{\sc III}}]15.6\mum}}{F_{[\rm{Ne\,{\sc II}}]12.8\mum}}
\end{equation}

We get a radiation field strength of $(9.1\pm 0.5)\times 10^{-20}\: $W~cm$^{-2}$, $(12.7\pm 0.9)\times 10^{-20}\: $W~cm$^{-2}$ and $(14.25\pm 0.5)\times 10^{-20}\: $W~cm$^{-2}$ for fields A, B, and C, respectively.

\subsection{[\silii] emission}

\begin{figure}
  \centering
  
\includegraphics[scale=0.5,angle=0]{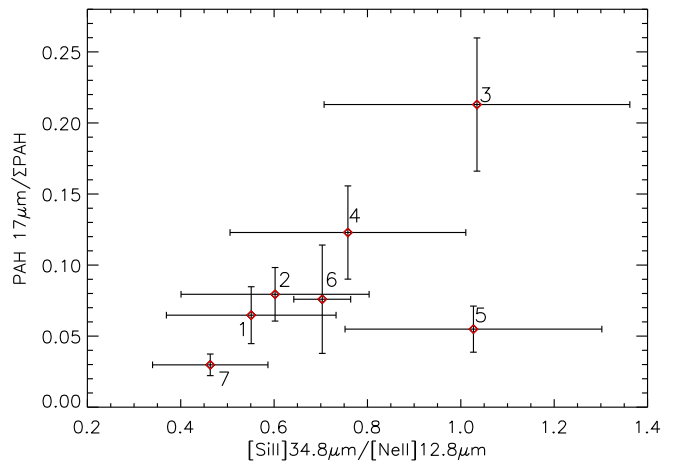}

  \caption{The [\silii]/[\neii] ratio plotted against the ratio of PAH$_{17\mum}$ strength to the strength of all the other PAH features, for our sources. Error bars in both quantities are shown.}

\label{fig:si_ii_pah}
\end{figure}

Most of the lines listed in Table \ref{tab:nebular} are from regions with ionized hydrogen gas. The exception is [\silii], which originates in a variety of environments including \hii\ regions, but also X-ray dominated regions \citep{Maloney96}, high density PDRs \citep{Kaufman06}, and regions of shocked gas, where heavy elements are returned to the gas phase. It is in general hard to pin down the physical mechanism for the emission of [\silii]. We investigate this in NGC~604 by looking at a possible correlation between the ratio [\silii]/[\neii] and the ratio of PAH emission at 17~\mum\ to all the other PAH features together. We plot the relation between these two ratios in Fig.~\ref{fig:si_ii_pah}. In \S \ref{sec:si_emission} we discuss this finding in the context of several possible scenarios for the emission of [\silii].

\subsection{H$_2$ emission}
Additional diagnostics on the physical conditions in NGC~604 come from the molecular hydrogen lines arising from pure rotational transitions at 12.27~\mum\ (0-0 S(2)) and 17.03~\mum\ (0-0 S(1)), which we detect (although marginally in the case of the 12.27~\mum\ line) in the hires modules of the IRS in fields A, B, and C. Using the same procedure as for the nebular lines, we fit Gaussian profiles to the H$_2$ rotational lines and estimate their strengths. In Table \ref{tab:h2_lines} we list the measured line strengths.

\begin{deluxetable}{ccc}
\tablecolumns{3}
\tablecaption{H$_2$ rotational lines}

\tablehead{
  \colhead{Source} &
  \colhead{H$_2$ S(2)12.27\mum} &
  \colhead{H$_2$ S(1)17.03\mum} \\
  \colhead{ } &
  \colhead{[$10^{-20}\times$W~cm$^{-2}$]} &
  \colhead{[$10^{-20}\times$W~cm$^{-2}$]} 
}

 \startdata

Field A &  0.19(0.15) & 0.20(0.01)  \\
Field B &   0.13(0.15) & 0.13(0.02) \\
Field C &  0.16(0.04) & 0.14(0.01)  \\
    
 \enddata

\label{tab:h2_lines}
\end{deluxetable}

H$_2$ temperatures can be calculated from the ratio of the two detected line strengths listed in Table \ref{tab:h2_lines}, using a general method \citep[see, for example the Appendix of][]{Brandl09}. This method holds under the following assumptions: \emph{(i)} the gas is in local thermodynamic equilibrium (LTE); \emph{(ii)} the hydrogen rotational lines are optically thin; \emph{(iii)} the critical densities for the lines are $< 10^3\: \rm{cm}^{-3}$; and \emph{(iv)} the two states of the H$_2$ molecule, ortho-H$_2$ and para-H$_2$, exist in a ratio of 3 to 1. From the Boltzmann statistics that describe the distribution of energy states for the H$_2$ molecule and the relation between the line strength and the H$_2$ column density, it can be shown that the excitation temperature is given by:

\begin{equation}
\label{h2_temp}
T_{\rm{ex}}=\frac{E_2-E_1}{k \ln \left(C\frac{\lambda_1}{\lambda_2}\frac{F_1}{F_2}\right)}
\end{equation}

where $E1$ and $E2$ are the energies of the involved levels, $k$ is the Boltzmann constant, $C$ is a constant dependent on the ratio of Einstein coefficients and statistical weights, $\lambda_1$ and $\lambda_2$ are the rest-frame wavelengths of the observed lines, and $F_1$ and $F_2$ their respective integrated line fluxes. Fields A and B have uncertainties close to 100\% in the H$_2$ S(2) flux (Table \ref{tab:h2_lines}), and hence we are unable to say anything conclusive about the gas temperature. Field C is comparatively better, and for this source we obtain $T_{\rm{ex}}$ between 200~K and 800~K.

\subsection{SED modeling}\label{sec:bayesian}
In MG11 we have presented a Bayesian fitting tool to study the infrared SEDs of star-forming systems. The tool is based on a grid of the Dopita \& Groves models (D\&G models hereafter), which are thoroughly described in a series of papers \citep{Dopita05, Dopita06b, Dopita06c, Groves08}. Here we briefly describe the tool and then apply it to the observed infrared SED of NGC~604 in \S \ref{sec:evolutionary}, including both the IRS spectrum and the PACS photometry, to derive statistically meaningful values for the following physical parameters of the region: cluster age and stellar mass, compactness, fraction of total mass contained in embedded objects and fraction of the luminosity arising in photon-dominated regions.

\subsubsection{Physics}

The D\&G models combine stellar synthesis, radiative transfer, dust physics, and self-consistent dynamical evolution of individual \hii\ regions to simulate the UV to sub-millimeter SEDs of star-forming regions, including thermal  emission from dust and PAHs, and nebular line emission. Using all the available spectral information, for a given metallicity ($Z$) and ambient interstellar pressure ($P_0/k$), the fitting tool computes probability distribution functions (PDFs) for key model parameters such as stellar cluster age ($t_{\rm{cl}}$), cluster mass ($M_{\rm{cl}}$),  compactness ($\mathcal{C}$), fraction of the total luminosity arising in the PDRs ($f_{\rm{PDR}}$), and the mass contribution from a component of young embedded objects that we model as Ultra Compact \hii\ Regions ($M_{\rm{emb}}$). 

We have defined these parameters in MG11. Of relevance here is the definition of $\mathcal{C}$, originally described in \citet{Groves08}. This parameter controls the form of the FIR continuum and the wavelength of the FIR peak. The shape of the SED at wavelengths longer than about 15~\mum\ is a function of the distribution of dust temperatures throughout a given region. Such distribution is a function of the total heating flux incident on the dust grains. For a spherical expanding nebula, the heating flux is calculated as $L_*/R_{\rm{HII}}^2$ at each time, where $L_*$ is the luminosity of the cluster and $R_{\rm{HII}}^2$ is the radius of the \hii\ region. Hence, models that preserve the run of dust temperature with time should also preserve the quantity $<L_*(t)>/<R_{\rm{HII}}(t)^2>$. The compactness parameter is proportional to this ratio, and hence sets the evolution dust temperature with time. Intuitively, for a given cluster luminosity, denser \hii\ regions will have smaller radii, and hence hotter dust.

\subsubsection{Priors}
Bayesian inference allows to include any previous evidence on the values of the model parameters in the calculation of the posterior PDF. Here we use flat, bounded priors for all the model parameters, with boundaries set by observational and theoretical studies of star-forming regions, as discussed in MG11. We assume solar metallicity and adopt a thermal pressure of the surrounding interstellar medium (ISM) of $P_0/k=10^5\: \rm{K}\: \rm{cm}^{-3}$. 

The assumption on ISM pressure is supported by measurements of far-IR line ratios from the ISO satellite on a sample of star-forming galaxies \citep{Malhotra01}. The metallicity of NGC~604 has been measured to be about half-solar \citep{Magrini07}. However, our experiments show that only the solar metallicity models in our grid are able to reproduce the observed ratio of PAH strength to continuum emission at 100~\mum\ in NGC~604. The discrepancy is most likely due to the specific PAH template used in the D\&G models, which is an empirical template based on observations of starburst galaxies. The choice of metallicity does not only affect the PAH emission strength. At far-IR wavelengths, the change in dust column and mechanical luminosity of the starburst with metallicity produces a slight shift and a broadening of the far-IR bump. While this implies an additional degeneracy with the compactness parameter, the additional errors introduced in the determination of $\mathcal{C}$ are smaller than our chosen step size for the compactness grid, which is 0.5.

\subsubsection{$\chi^2$ Weighting}
In MG11 we have included a discussion of the observational uncertainties involved in measuring the IRS fluxes, which include absolute and relative flux calibrations and systematic errors due to specific observing conditions. Based on that discussion, for fitting purposes we have adopted a uniform uncertainty of 10\% across the IRS wavelengths, which also implies a uniform weighting for all data points in the fitting. 

The $\chi^2$ minimization procedure described in MG11 has been slightly modified here to ensure that the $\chi^2$ minimization is not dominated by the IRS range, which has many more resolution elements, as compared to only two data points at 100~\mum\ and 160~\mum\ from PACS. Each bin contributes to the $\chi^2$ with a weight that is proportional to the bin size in the logarithmic wavelength space. The bin size is set by the resolution of the models in the IRS range and by the wavelength separation between data points for the PACS data. This results in a weighting function that increases uniformly in the IRS range so that the bins at the short wavelength end (around 5~\mum) have about half the weight of those at 35~\mum\, and about 0.25 times the weight of the PACS bins. In section \S \ref{sec:evolutionary} we apply this tool to the observed SED of NGC~604.

\subsubsection{Color Correction of PACS photometry}
The D\&G models compute the monochromatic flux densities for each wavelength bin. However, the flux densities measured by the PACS filters at 100~\mum\ and 160~\mum\ are not monochromatic. They are the integrated flux densities over certain wavelength ranges, as modulated by the filter response function. Before we compare the model fluxes to the observed PACS photometry, it is necessary to evaluate the errors introduced by this difference. To do so, we adopt the method described in \citet{DaCunha08} to predict the flux density of a source when observed using a given filter. According to their approach, the flux density for any filter is:

\begin{equation}
\label{color_corr}
F_{\nu}^{\lambda_0}=\frac{\lambda_0^2}{c}C_{\nu_0}\frac{\int \rm{d}\lambda\: F_{\lambda}\: \lambda R_{\lambda}}{\int \rm{d}\lambda\: C_{\lambda}\: \lambda\: R_{\lambda}}
\end{equation}

where

\begin{equation}
\label{lambda_0}
\lambda_0=\frac{\int \rm{d}\lambda\: \lambda\: R_{\lambda}}{\int \rm{d}\lambda\: R_{\lambda}}
\end{equation}

is the effective wavelength of the filter response $R_{\lambda}$, $c$ is the speed of light, and $C_{\lambda}$ is a calibration spectrum that depends on the photometric system used for the calibration. In the case of the PACS photometer, the calibration spectrum is constant ($C_{\lambda}\lambda=\rm{constant}$), which simplifies the calculations. Using this method, and the filter response curves available at the Herschel Science Center, we have computed effective wavelengths for the PACS filters at $L_0^{100}=102.6~\mum$ and $L_0^{160}=167.2~\mum$. Furthermore, using Eq. \ref{color_corr}, we have estimated by how much the PACS fluxes, as predicted by our models, should be corrected once the color correction is applied. For our best fit model, the differences between the monochromatic model fluxes and those that would be measured by PACS are $<1\%$ for the PACS green filter (100~\mum) and $\sim 9\%$ for the PACS red filter (160~\mum). These uncertainties are within the 10\% observational errors.

\section{Discussion}
\label{sec:discussion}

\subsection{Notable Sources}
\label{sec:notable}

The most striking morphological characteristic of the sources labelled 1-7 in Fig.~\ref{fig:n604_rgb_map_first} is that most of them (sources 1-5) are well defined, individual infrared-bright knots. Sources 1, 2, 7, and 4 are, in that order, the strongest sites of PAH emission, and also the sites with warmer dust continuum, as shown in Tables \ref{tab:cont_slope} and \ref{tab:pah_features}. This implies warmer dust temperatures near sources with stronger 8~\mum\ emission, where most of the MYSO candidates have been identified by \citet{Farina12} using NIR photometry.

\subsubsection{Source 2}
This source, located in the SE lobe, is bright in all IRAC bands but has no optical, H$\alpha$ or FUV counterpart. In addition, it has one of the strongest silicate absorptions at 10~\mum\, as shown in Fig.~\ref{fig:all_spectra}, and it is only about $5^{\prime\prime}$ from the peak of one of the CO clouds first reported in \citet{Wilson92}. Using HST data, \citet{Maiz_Apellaniz04} derived an extinction map for the region. They found a strong peak of extinction at the location of source 2, which is consistent with its relatively strong silicate absorption. Compared with all the other sources, nebular emission towards this source is weak (Table \ref{tab:nebular}). Its high optical extinction, bright PAH emission, and spatial coincidence with a reservoir of molecular gas  makes of source 2 a very good candidate for a site of embedded star formation. 

\subsubsection{Source 5}
Source 5 shows the coldest dust temperature among our sources, as traced by the $\rm{F}_{15\mum}/\rm{F}_{30\mum}$ ratios listed in Table \ref{tab:cont_slope}. Its [\siv]10.4\mum/[\siii]18.7\mum\ and [\neiii]15.5\mum/[\neii]12.8\mum\ line ratios derived from Table \ref{tab:nebular} are both the lowest among our sources and the PAH features listed in Table \ref{tab:pah_features} are weaker towards this source, compared with all the other well defined IR knots. The combination of cold dust, low ionization state, soft radiation field (as traced by the line ratios), and weak emission from PDRs are indicative of a more evolved stage for this particular source. 

Source 5 is also brighter than all the other sources at 3.6~\mum. This band traces the 3.3~\mum\ PAH feature, and photospheric emission from stellar populations older than 10~Myr. Since the continuum PAH emission is weak towards this source, the most plausible explanation for the excess at 3.6~\mum\ is again the presence of stars older than 10~Myr. In fact, this source coincides with the location of one of the WR stars identified by \citet{Hunter96}, and is located near the boundary between the active star-forming western and quiescent and older eastern hemisphere of NGC~604, as described by \citet{Tullmann08}.

\subsubsection{Source 7, Source 4, and Source 1}
\label{sec:sources741}
Source 7, located in the NW infrared lobe, shows the strongest nebular line emission among our sources (Table \ref{tab:nebular}). This source is very close to a bright ridge of photospheric optical emission from a nearby group of young stars (see Fig.~\ref{fig:n604_rgb_map_first}). The higher electron density near the NW lobe is consistent with the strong nebular lines measured near source 7 and with the enhanced H$\alpha$ emission observed in this same area of NGC~604 \citep[see, for example, Fig.~2 in][]{Relano09}, and implies a higher ionization state. This source is most likely the location of the youngest main sequence stars of the cluster. This is supported by our measurements of the radiation field strength in the neighboring field C, where we have measured a stronger radiation field using Eq.~\ref{field_strength}, relative to the other fields. In fact, UV spectroscopy using HST's \emph{Space Telescope Imaging Spectrograph} has revealed that field C coincides with the location of a young, luminous O7 star, most likely one of the most massive stars in NGC~604 \citep{Bruhweiler03}.

The [\siv]/[\neii] ratio map of Fig.~\ref{fig:siv_over_nii} peaks within the projected area of source 4, which has a relatively harder radiation field, where we have also measured a warm dust component from the spectral continuum slope (Table \ref{tab:cont_slope}). Another location with relatively hard radiation field is source 1. The \citet{Relano09} study of the region reveals enhanced H$\alpha$ emission in sources 1, 4, and 7. However, the eastern part of NGC~604, where sources 1 and 4 are located, has a higher optical extinction traced by the Balmer optical thickness ($\tau_{\rm{Bal}}\sim 1.2$), as compared with the surrounding average extinction ($\tau_{\rm{Bal}}= 0.2-0.3$) \citep{Maiz_Apellaniz04}. The [\siv]10.5\mum\ falls on the broad silicate absorption feature near 10~\mum, and this leads to an underestimation of the [\siv]/[\neii] ratio in regions of high extinction. Although this does not affect our general picture about the radiation hardness. Larger [\siv]/[\neii] ratios and hence younger ages might be expected in sources 1 and 4 from extinction corrected [\siv]/[\neii] ratios. 

Our results are consistent with the optical spectroscopy of the region presented in \citet{Maiz_Apellaniz04} in terms of the high density and excitation derived near sources 1, 4, and 7, as well as field C, and point to the fact that these are the regions where the youngest, most massive stars are located within NGC~604. Finally, the [\siv]10.5\mum/[\neii]12.8\mum\ ratio is sensitive not only to the cluster age, but also to the gas density. However, at the relatively low electron densities measured in the region ($\log n_e \approx 1.5-2.5\: \rm{cm}^{-3}$), we do not expect the measured line ratios to be significantly affected.

\subsection{The evolutionary status of NGC~604}
\label{sec:evolutionary}

\begin{figure}
  \centering
  
\includegraphics[scale=0.5,angle=90]{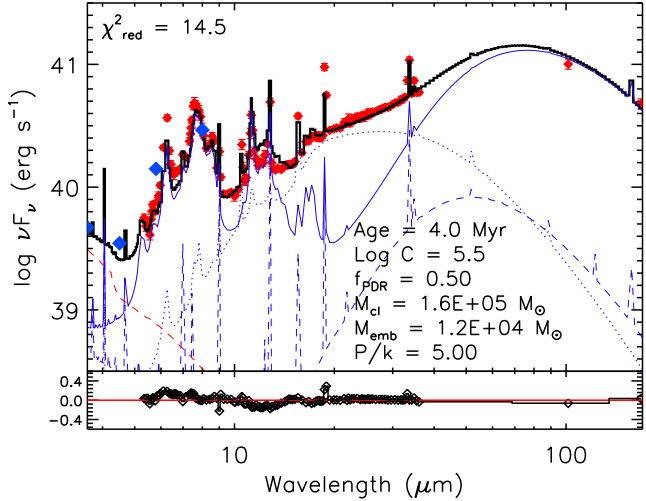}

  \caption{Best fit model to the integrated SED of NGC~604.  The IRS and PACS data are shown as red diamonds, and the IRAC photometry as blue diamonds. The solid line is the best fit model. Also shown are the contributions from pure \hii\ region emission (dashed blue), PDR (solid blue), embedded population (dotted blue), and photospheric emission from stars older than 10~Myr (dashed red). Residuals are shown in the bottom panel.}

\label{fig:best_fit_int}
\end{figure}

\begin{figure*}[h!t]
  \centering
  \subfigure[]{
\includegraphics[scale=0.23,angle=90]{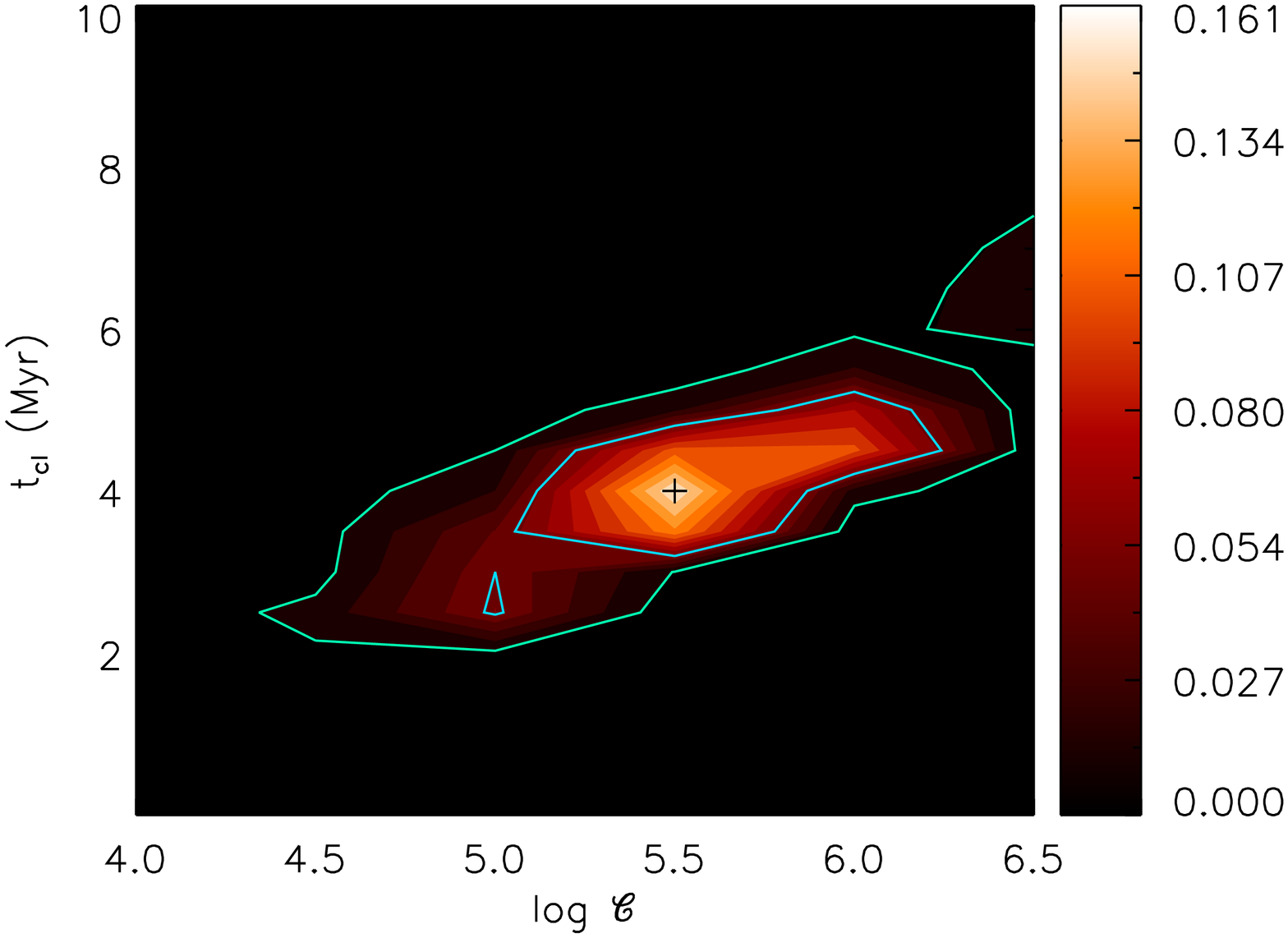}
\label{fig:pdfs_a}
}
  \subfigure[]{
\includegraphics[scale=0.23,angle=90]{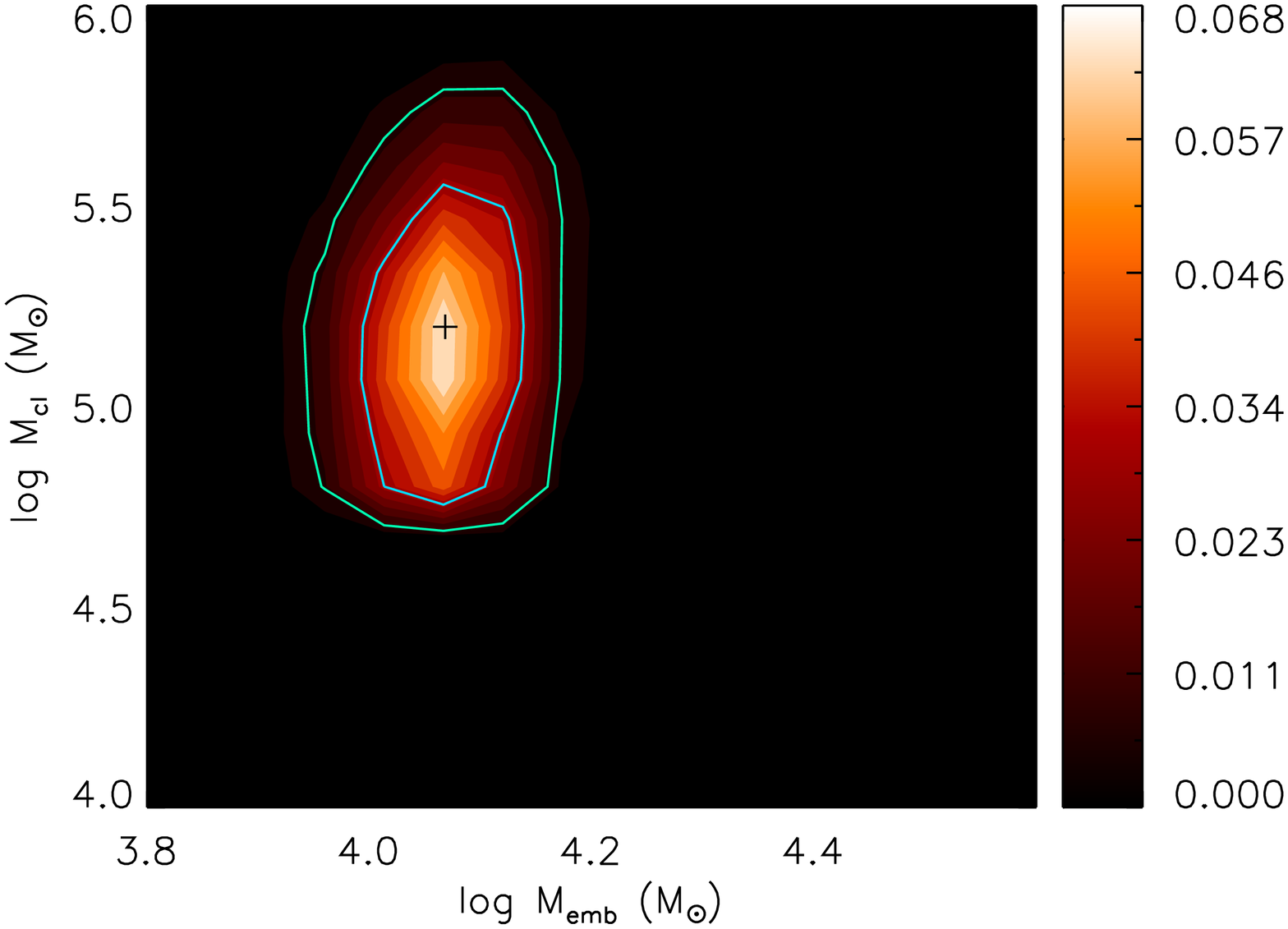}
\label{fig:pdfs_b}
}
\subfigure[]{
\includegraphics[scale=0.23,angle=90]{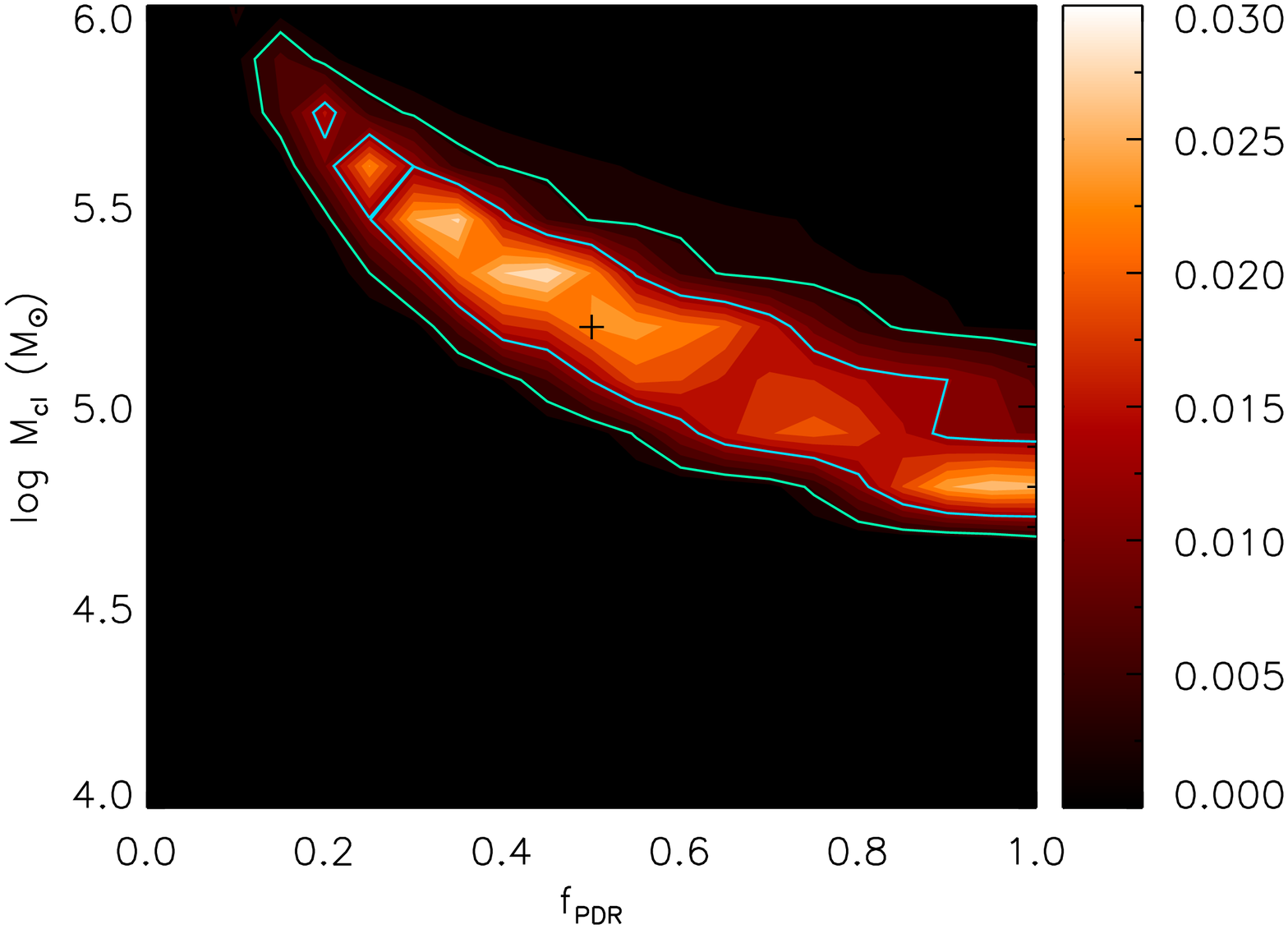}
\label{fig:pdfs_c}
}
 
  \caption{Two-dimensional PDFs for selected pairs of parameters covering the totality of the parameter space. The pairs shown are: \emph{(a)} $\log \mathcal{C}$ - $t_{\rm{cl}}$, \emph{(b)} $\log \rm{M}_{\rm{emb}}$ - $\log \rm{M}_{\rm{cl}}$, and \emph{(c)} $f_{\rm{PDR}}$ - $\log \rm{M}_{\rm{cl}}$. The color code indicates the normalized probability. The cross symbols mark the best-fit values while the color contours indicate the 1-$\sigma$ (blue) and 90\% (green) confidence levels.}

\label{fig:pdfs}

\end{figure*}

Fig.~\ref{fig:best_fit_int} shows the best fit to the observed integrated SED of NGC~604 (IRS + PACS) using the Bayesian tool described in \S \ref{sec:bayesian}. Also, in Fig.~\ref{fig:pdfs} we show the resulting PDFs, covering the totality of our parameter space. In Table \ref{tab:results_int} we list the associated best fit parameters and uncertainties, with the values for $t_{\rm{cl}}$ and $M_{\rm{cl}}$ determined independently by \citet{Eldridge11} using several methods, including optical SED fitting. We also include $f_{\rm{emb}}$, which is the ratio of mass contained in young embedded objects ($M_{\rm{emb}}$) to stellar mass in the cluster ($M_{\rm{cl}}$). This Bayesian fit, as well as the other spectral features discussed above, provide a wealth of information about the physics and the evolutionary stage of NGC~604.

\begin{deluxetable}{ccc}
\tablecolumns{3}
\tablecaption{Best fit to the integrated spectrum of NGC~604\tablenotemark{a}}

\tablehead{
  \colhead{Parameter} &
  \colhead{Best fit value} &
  \colhead{\citet{Eldridge11}}
  }

 \startdata

  $t_{\rm{cl}}$ (Myr)                                    &  $4.0_{-1.0}^{+1.0}$     &   $>4$   \\ [3pt]
  $\log \mathcal{C}$                                    &   $5.5_{-0.5}^{+0.5}$    &      \\ [3pt]
 $f_{\rm{PDR}}$                                          &   $0.5_{-0.2}^{+0.3}$    &       \\ [3pt]
 $M_{\rm{cl}}$ ($10^5\: \rm{M}_{\astrosun}$)               &   $1.6_{-1.0}^{+1.6}$    &  $(3.8 \pm 0.6)$    \\ [3pt]
 $M_{\rm{emb}}$ ($10^4\: \rm{M}_{\astrosun}$)              &   $1.2_{-0.1}^{+1.3}$    &       \\ [3pt]
 $f_{\rm{emb}}$                                          &   $0.075$              &       \\ [3pt]

 \enddata
\tablenotetext{a}{Parameters listed are cluster age ($t_{\rm{cl}}$), compactness ($\log \mathcal{C}$), fraction of total luminosity arising in PDR regions ($f_{\rm{PDR}}$), stellar mass of the cluster ($M_{\rm{cl}}$), mass contained in embedded objects ($M_{\rm{emb}}$), and ratio of embedded to stellar mass ($f_{\rm{emb}}$).}

\label{tab:results_int}
\end{deluxetable}

\subsubsection{Dust temperature and compactness}

The thermal continuum, PAH emission, and line emission are well reproduced by the best fit.  The combined IRS and PACS data are consistent with a far-IR SED peaking at approximately 70~\mum, which indicates an effective dust temperature of $\sim 40\: $K. Using multi-wavelength photometry of a sample of \hii\ regions in the Magellanic Clouds, \citet{Lawton10} show that the infrared SEDs of most of these regions peak at 70~\mum. Our result indicates that this average dust temperature is not unique of ``local'' \hii\ regions, but also applies to at least one relatively more distant region.

We have stated that the compactness parameter controls the incident heating flux $L_*/R_{\rm{HII}}$ as a function of age of the \hii\ region. From our given set of best fit parameters it is then possible to derive an incident heating flux \citep[see Fig.~3 in][]{Groves08}. We obtain an incident flux of $\log (L_*/R_{\rm{HII}}) = 0.65\: \rm{erg}\: \rm{s}^{-1}\: \rm{cm}^{-2}$ for our best fit values of compactness and age. The bolometric luminosity of the cluster can be obtained from integration of the best fit SED, and gives approximately $6.0\times 10^{41}\: \rm{erg}\: \rm{s}^{-1}$, which allows us to solve for the effective radius of the \hii\ region. We obtain $R_{\rm{HII}}\sim 120\: $pc. This value corresponds to the scale of the larger filaments observed in NGC~604 (Fig.~\ref{fig:n604_rgb_map_first}). This agreement between a model-derived quantity and a measurable observable demonstrates the capabilities of our Bayesian approach to obtain valuable physical information when applied to unresolved star forming regions.

\subsubsection{Stellar mass}
The total stellar mass of $1.6_{-1.0}^{+1.6}\times 10^5\: \rm{M}_{\odot}$ that we estimate with the SED fitting is in agreement, within the uncertainties, with the mass derived by \citet{Eldridge11}. Besides, within the uncertainties, this mass is similar to the mass derived for 30 Doradus in MG11, which equals $0.63_{-0.23}^{+0.37}\times 10^5\: \rm{M}_{\odot}$. 

In general, previous studies have found that the total stellar mass in 30 Doradus is larger than in NGC~604. We propose two reasons to explain the fact that the value of our derived mass for NGC~604 is larger than the value of the mass derived for 30 Doradus. First, the physical projected area covered in our NGC~604 map is relatively larger than the projected area of the 30 Doradus map, and hence there is some missing emission from the periphery in the latter case that does not enter the SED. Second, we observe in Fig.~\ref{fig:pdfs_c} a degeneracy between the cluster mass and the fraction of the total luminosity from PDRs that covers a range of masses of $\sim 0.5\: $dex. A similar degeneracy was found for 30 Doradus in MG11. We have argued that the reason for this degeneracy is the fact that both the PDR and the pure \hii\ region spectra contribute to the dust thermal continuum. An increase in PAH from emission is accompanied by an increase in the overall infrared luminosity. This also renders the additional far-infrared data insufficient to solve this degeneracy. Optical wavelengths data, where the emission from PDR is significantly different from that of an \hii\ region, are needed to settle this issue. Nevertheless, although our result is subject to this $M_{\rm{cl}}-f_{\rm{PDR}}$ degeneracy, a comparison between our derived masses and the masses derived for 30 Doradus and NGC~604 in \citet{Hunter95} and \citet{Eldridge11} respectively, implies that our Bayesian fitting tool can find solutions that are in reasonable agreement with published literature values.

\begin{figure*}
  \centering
  \subfigure[]{

\includegraphics[scale=0.45,angle=90]{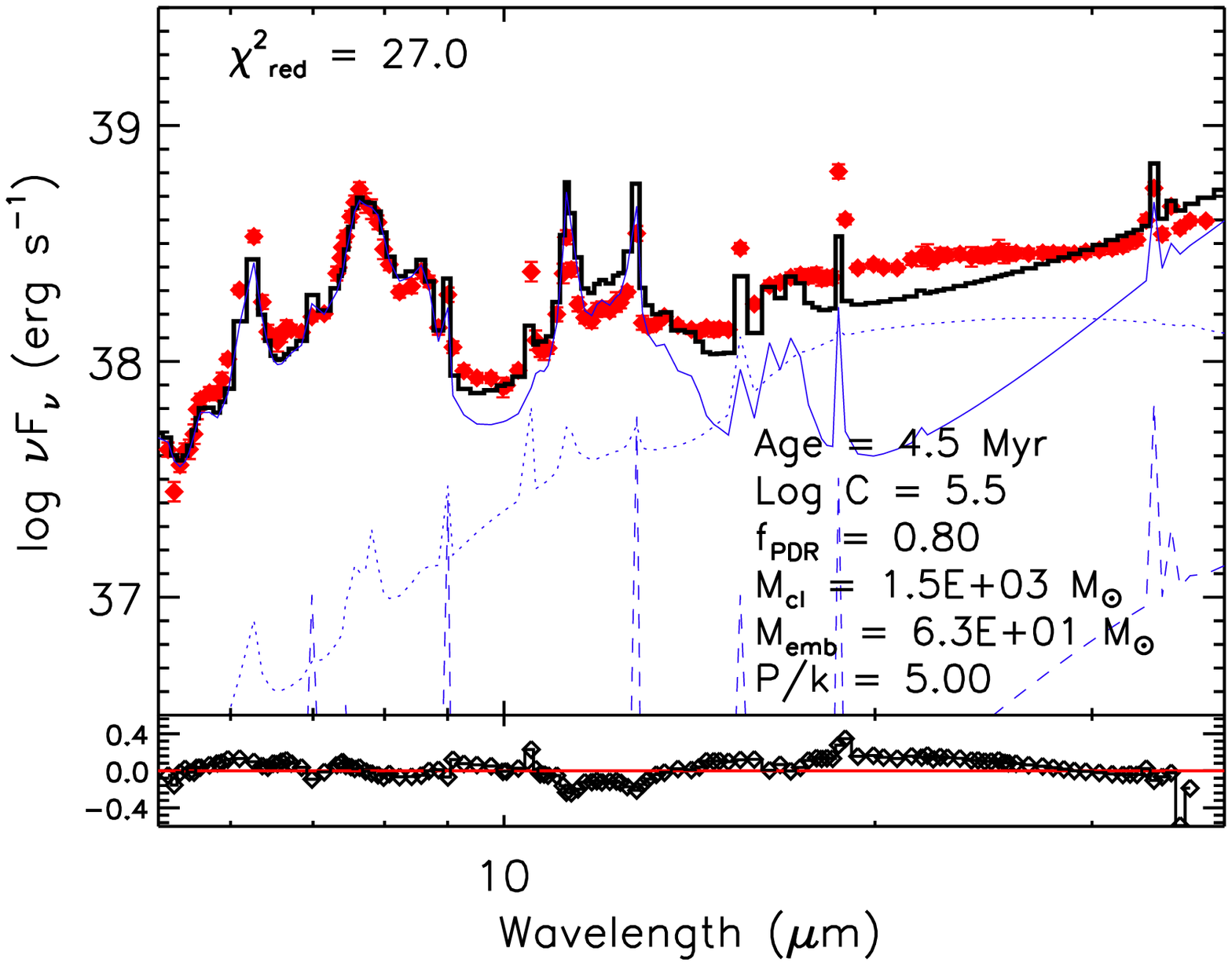}

}
  \subfigure[]{
\includegraphics[scale=0.45,angle=90]{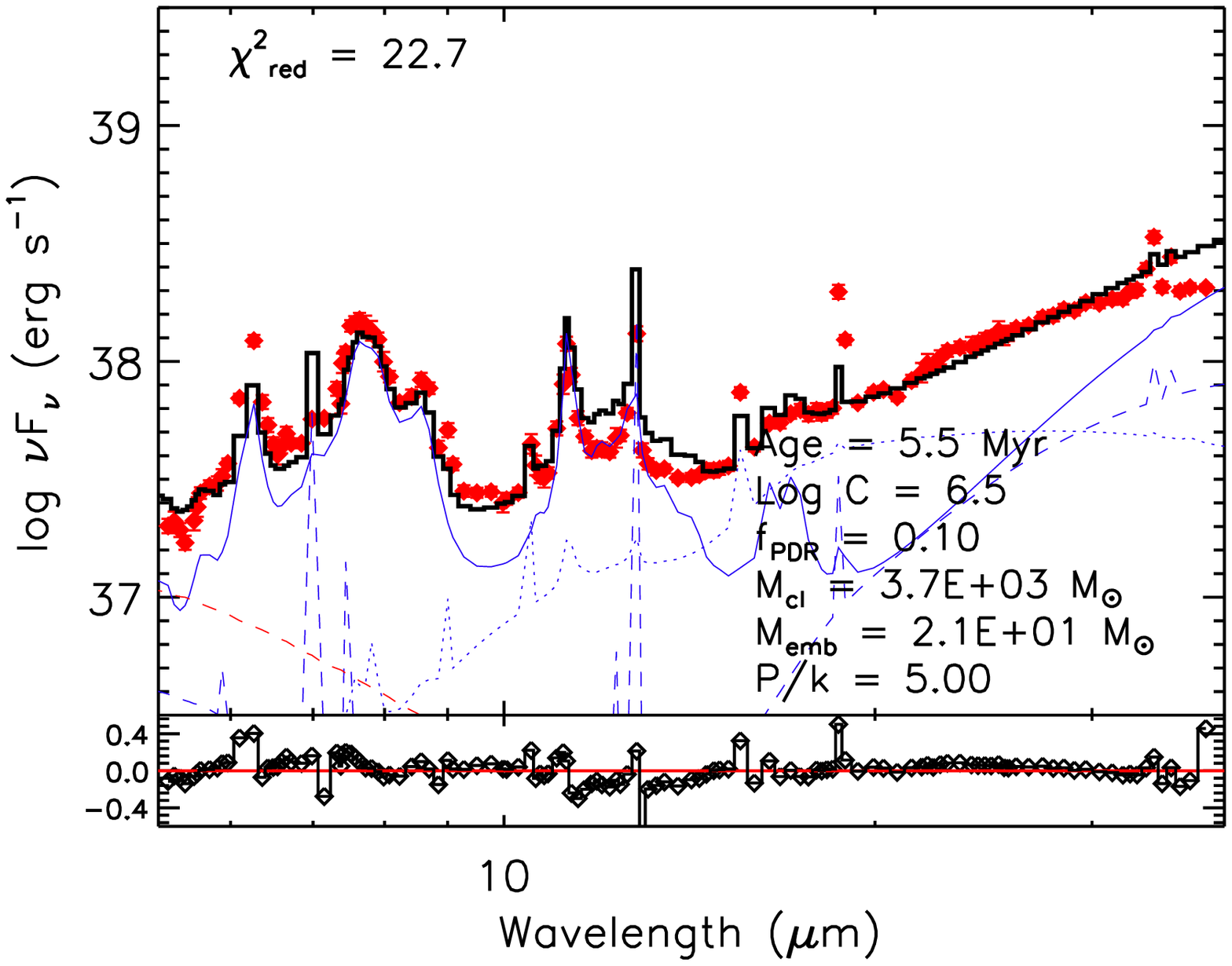}

}

  \caption{SED fits to the spectra of source 1 (a) and 5 (b). Color codes and lines are the same as in Fig.~\ref{fig:best_fit_int}. The SED luminosities are consistent with masses of a few times $10^3~\rm{M}_{\astrosun}$.}

\label{fig:fit_sources}
\end{figure*}

A brief discussion on the mass of the individual infrared sources in Fig.~\ref{fig:n604_rgb_map_first_b} is relevant here. Although the SED models are not intended to reproduce ensembles of embedded objects but entire \hii\ regions, we fit their spectra using our tool to estimate their individual masses. In Fig.~\ref{fig:fit_sources} we show the fit to the IRS spectra of two representative sources that significantly differ in their IRAC colors: sources 1 and 5. Note that the masses of these sources are of the order of a few times $10^3~\rm{M}_{\astrosun}$. This is the case for all the other sources, which are similarly luminous. Although we do not attempt a detailed interpretation of the fitted parameters, because we are aware of the limits of our models, it is qualitatively interesting that we derive a higher PDR contribution and a higher fraction of embedded mass for source 1 than for source 5, which is consistent with the latter being in a region dominated by stars older than 10~Myr (see \S \ref{sec:notable}).

\subsubsection{Age of the region}

Both the \emph{instantaneous burst} approach that leads to a single, coeval stellar population and the \emph{continuum star formation rate} approach that leads to a uniform distribution of stellar ages, are opposite idealizations of the star formation history in a particular starburst region. In fact, rather than showing one of these limiting situations, it is often found that several star formation events in a particular region lead to different stellar populations of different ages. NGC~604 is not an exception to this rule. \citet{Eldridge11} argue that the populations of Wolf-Rayet (WR) stars and red supergiants (RSG) in NGC~604 belong to different formation events, with estimated ages of $3.2\pm 1.0\: $Myr and $12.4\pm 2.1\: $Myr, respectively. A similar situation is observed in other giant \hii\ regions such as 30 Doradus. 

Henceforth, when we refer here to the age of the region what we mean is the age of the dominating stellar population, which is usually composed of the youngest, most massive main sequence stars in these systems. As we have already discussed, most of the infrared continuum and line emission up to wavelengths of 100~\mum, arises from the interaction of the strong radiation field from young ($<10\: $Myr) stars and the surrounding ISM. Only at longer wavelengths, diffuse dust heated by older stellar populations starts to dominate the SED. In the following we discuss how our best fit parameters for NGC~604 help us constrain the age of its dominating stellar population.

The fraction of total mass contained in embedded objects that we measure here for NGC~604 ($f_{\rm{emb}}\sim 0.08$) is smaller than that measured for 30 Doradus in MG11 ($f_{\rm{emb}}\sim 0.53$). If we interpret this result in terms of recent star formation taking place in these two regions, this suggests that 30 Doradus has a larger relative amount of mass in embedded objects, but this conclusion should be taken with caution, since other dust heating mechanisms can play a role. Additionally, the age of $\sim 4\pm 1~\rm{Myr}$ derived with our fitting method is consistent with the study of the line ratios that we presented in \S \ref{sec:hardness} and agrees well with the studies by \citet{Eldridge11} and \cite{Hunter96}. In MG11 using the same method, for 30 Doradus we have obtained an age of $3\pm 1.5~\rm{Myr}$. Our results suggest that the dominating stellar population in each of these two giant HII regions (i.e., the one that provides most of the ionizing photons) are consistent with slightly different model ages when variations in all model parameters are considered

\subsubsection{PDR content}

Although the fraction of total luminosity arising in PDRs is degenerate with the total cluster mass, as shown in Fig.~\ref{fig:pdfs_c}, our result from SED fitting suggests that at least half of the energy output from NGC~604 is generated in PDRs, the other half arising in ionized regions closer to the cluster stars. The presence of a luminous PDR in NGC~604 is consistent with the findings of \citet{Heiner09}, who find considerably high amounts of atomic hydrogen ($N_{\rm{HI}}\sim 2.0 \times10^{21}\: \rm{cm}^{-2}$), and associate them with the photodissociation of $H_2$ in a dense ($n\sim 500\: \rm{cm}^{-3}$) PDR.

\subsubsection{Molecular hydrogen in NGC~604}

The $\rm{H}_2$ excitation temperature near field C ($T_{\rm{ex}}\sim 200~\rm{K}-800~\rm{K}$), calculated using Eq.~\ref{h2_temp} is indicative of the average temperature of the warm gas in NGC~604. Additionally, for a given value of the molecule angular momentum, the total number of hydrogen molecules is proportional to the strength of the line. 

Our results on the H$_2$ S(1) line, which has smaller uncertainties (Table \ref{tab:h2_lines}), indicate that warm molecular hydrogen is present in fields A, B, and C and slightly more abundant towards field A, which has the strongest H$_2$ lines. This result is consistent with the findings of \citet{Tosaki07}, who report the detection of an arc-like distribution of warm and dense molecular gas that extends across coincident our fields A, B and C, and associate it with gas compression by the expanding \hii\ region and sequential star formation in NGC~604. The center of Field A is, in projection, about $5^{\prime\prime}$ (or 20~pc) away from the peak of CO emission first reported in \citet{Wilson92} and confirmed to be a dense molecular cloud in \citet{Miura10}, although given the extension of this cloud (about 30~pc), we can consider it as coincident with Field A. We cannot confirm a physical association of the molecular cloud with our infrared sources, since we only see a projected spatial association. However, based on our results and the mentioned previous work, it becomes evident that both dense, molecular gas as well as warm molecular hydrogen is detected in the line of sight towards Field A.

\subsubsection{Selective destruction of small PAHs}\label{sec:PAH_proc}

\begin{figure}
  \centering
  
\includegraphics[scale=0.43,angle=0]{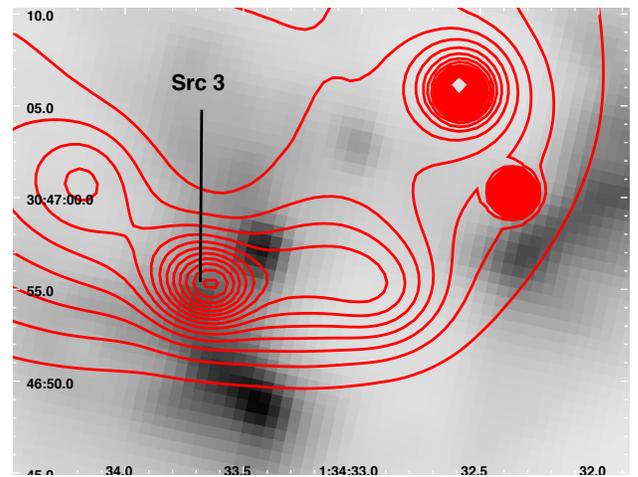}

  \caption{Soft X-ray emission in NGC~604. The black and white image is the IRAC 8~\mum\ emission. The red contours correspond to the ACIS image. Source 3 coincides with a local peak of soft X-ray emission with a flux of $3.70\times 10^{-9}\: \rm{erg}\: \rm{cm}^{-2}\: \rm{s}^-1$ (see Table \ref{tab:pah_ratios}).}

\label{fig:soft_x}
\end{figure}

The 17~\mum\ PAH feature is generally associated with out-of-band bending modes of large PAH molecules, containing $\approx 2000$ C-atoms \citep{Kerckhoven00}. Table \ref{tab:pah_ratios} shows that the relative strength of this feature exhibits an increasing trend with the X-ray flux. If we adopt the PAH size argument, this implies that it is the ratio of large to small PAH molecules that scales with the X-ray field. In fact, Fig.~\ref{fig:soft_x} shows that source 3, which shows the largest enhancement of the 17~\mum\ PAH feature among our sources, coincides in projection with a peak of soft X-ray emission. Soft X-ray emission from massive stars can be associated with shocked stellar winds or magnetically confined gas near the wind base, near the stellar coronae \citep{Cassinelli83}. 

The dissociation of PAH molecules by X-rays has been discussed in \citet{Micelotta10}. They argue that not all X-ray photon absorptions by PAH molecules lead to photodissociation and estimate that after a second electron has been ejected by the PAH molecule via the Auger effect, the molecule is left with an internal energy of 14-35~eV, enough to dissociate small (50 C-atoms) PAHs, but possibly insufficient to dissociate larger molecules. They leave the question of large PAH survival open. Our observations suggest that, at least in the particular case of source 3, X-ray emission from massive stars may be responsible for the dissociation of small PAH molecules, creating an enhancement of the 17~\mum\ feature. Similar relative enhancements of the 17~\mum\ feature have been observed in regions of hard radiation fields in galaxies, such as the vicinity of AGNs \citep{Smith07, Odowd09}. It is worth mentioning here that a MYSO candidate has been identified by \citep{Farina12} less than $1^{\prime\prime}$ away from source 3. Also, \citet{Barba09} report the presence of at least 2 red supergiants (RSGs) coincident with the location of this source. If other RSGs have already exploded as supernovae near this location, this could provide part of the X-ray enhancement observed.

\subsubsection{The origin of the [\silii] emission}\label{sec:si_emission}

Fig.~\ref{fig:si_ii_pah} is suggestive of a correlation between the enhancement of [\silii] and the enhancement of 17~\mum\ PAH emission at scales of tens of parsecs. Despite the relatively large uncertainties in both the PAH ratio and the [\silii]/[\neii] ratio, the figure indicates clearly that the enhancement of the PAH 17~\mum\ feature scales with the relative strength of the [\silii] emission, with the exception of source 5, which lies outside this correlation. This result suggests a common underlying physical mechanism for the enhancement of both features. We have shown in Table \ref{tab:pah_ratios} that the regions of strongest 17~\mum\ PAH emission also correspond to regions of enhanced X-ray emission. Based on the triple correlation between PAH$_{17\mum}$ emission, [\silii] emission, and soft X-ray flux, we argue that absorption of ambient X-rays may be linked to both the destruction of small PAH molecules and the excitation of [\silii], and produces the correlation observed in Fig.~\ref{fig:si_ii_pah}.

As we have pointed out, source 5 is an outlier of this correlation: it shows a relatively large [\silii]/[\neii] ratio, but very moderate PAH$_{17\mum}/\Sigma PAH$ ratio. It also lies in a region of low X-ray flux, near the edge between the western, active star-forming hemisphere of NGC~604 and the quiescent eastern region where most of the Wolf-Rayet stars are located \citep{Drissen08}. In fact it coincides with the location of one of these WR stars \citep{Hunter96}. In this part of NGC~604, the relatively high X-ray luminosity is most likely powered by the shocked gas resulting from older dynamical processes. Based on these facts, we speculate that the [\silii]/[\neii] ratio for this source is enhanced due to the presence of shocked gas rather than X-ray induced. Although no supernova remnants (SNRs) have been identified in the eastern hemisphere, this is not surprising, since such objects expanding into a low-density gas are hard to detect \citep{Chu90}.

\subsection{Ongoing and Triggered Star Formation in NGC~604}

Several authors have gathered evidence that suggest the presence of ongoing star formation in NGC~604 \citep{Maiz_Apellaniz04, Tosaki07, Relano09, Farina12}. Nevertheless, none of these authors have provided spectroscopic confirmation for such very recent star formation activity in the region. Our study of the sub-condensations whose spectra are shown in Fig.~\ref{fig:all_spectra} allows us to quantify the amount of embedded star formation happening in NGC~604.

In \S \ref{sec:notable} we have discussed the IR-bright sub-condensations that we detect in the IRAC maps. At the distance of NGC~604 it is unlikely that they are individual massive young stellar objects. Their SED luminosities are consistent with stellar masses of a few times $10^3\: \rm{M}_{\astrosun}$ (Fig.~\ref{fig:fit_sources}). The exact values depend on the individual SED fits and the uncertainty in the determination of the stellar mass (Fig.~\ref{fig:pdfs_b}). This implies that all together they account for between 5\% and 20\% percent of the total stellar mass in NGC~604. We have shown in Fig.~\ref{fig:itera} that the line ratios measured towards these clumps, are consistent with ages of between 4~Myr and 4.5~Myr. Although the timescales for the dissipation of dusty envelopes around star-forming clusters can be as long as 10~Myr in regions of high density, the low electron density that we have inferred in \S \ref{sec:e_density} implies that NGC~604 is a rather diffuse region. Consequently, it is unlikely that our infrared clumps are evolved clusters. The presence of these young, massive, IR-bright clusters, some of which have deep silicate absorption features, is a clear indication of significant massive star formation currently taking place in NGC~604.

Additional evidence that supports this scenario comes from our SED fitting analysis. Fig.~\ref{fig:best_fit_int} shows that a warm component of dust is necessary to fit the spectrum of NGC~604 between 15~\mum\ and 30~\mum. This component (dotted blue line in Fig.~\ref{fig:best_fit_int}) arises from MYSO (that in the D\&G frame we model as Ultra Compact \hii\ Regions), and corresponds to dust temperatures of about 300~K. Although the required fraction of mass in embedded objects is considerably smaller than that derived for 30 Doradus in MG11, it suggests the presence of a considerable amount of embedded star formation in the region. The total embedded mass derived is in fact comparable to the sum up of the individual clump masses, and equals approximately 8\% of the total cluster mass, as can be calculated from Table \ref{tab:results_int}.

In an evolutionary context, sequential star formation in NGC~604 is a plausible scenario to explain our observations. This idea has been explored by other authors before. Based on submillimeter observations of the CO ($J=3-2$)/CO ($J=1-0$) ratio, \citet{Tosaki07} report the existence of a dense ridge of molecular gas that surrounds the main cluster in NGC~604 and extends in the SE-NW direction, closely following the location of our bright IR lobes. To explain their results, they adopt an scenario in which the compression of molecular gas by the mechanical input from the main cluster (the first generation of stars) has triggered a second generation of stars near the NW infrared lobe. The strong radiation field that we observe close to field C and source 7, where main sequence stars are clearly observed, supports the existence of this second generation of highly ionizing stars. Furthermore, the results discussed in this subsection indicate that massive star formation is currently taking place within the SE lobe of NGC~604, even further away from the main cluster, where we have identified the IR-bright sub-condensations of Fig.~\ref{fig:n604_rgb_map_first}, some of which are heavily enshrouded by dust (e.g., source 2).

\section{Conclusions}
\label{sec:conclusion}
We have investigated the physical conditions and quantified the amount of ongoing massive star formation in the star forming region NGC~604. We used a combination of observational and modeling tools, including infrared spectrophotometry and Bayesian SED fitting of Spitzer and Herschel data. Here are our main findings:

\begin{enumerate}

\item We have identified several individual bright infrared sources along the luminous PDRs that surround the ionized gas in NGC~604. These sources are about $15\: $pc in diameter and have luminosity weighted masses between $10^3\: \rm{M}_{\astrosun}$ and $10^4\: \rm{M}_{\astrosun}$. 

\item The deep 10~\mum\ silicate absorption feature, mid-IR continuum slope, and atomic line ratios towards some of these sources indicate that they are young embedded systems, and most likely the sites of ongoing massive star formation in NGC~604. Some of them (i.e. source 2) are also associated with gas reservoirs as traced by CO maps. This is in agreement with previous studies of the region.

\item A scenario of ongoing massive star formation in NGC~604 is supported by Bayesian fitting of the integrated spectrum (lines+continuum) of the region constructed from \emph{Spitzer}-IRS and \emph{Herschel}-PACS observations. Our results indicate that embedded star formation can account for up to 8\% of the total stellar mass in NGC~604. 

\item The spectral fitting also implies an age of $4.0\pm 1.0\: $Myr for the region and a total stellar mass of $\sim 1.6^{+1.6}_{-1.0}\times 10^5\: \rm{M}_{\astrosun}$. These results are in agreement with independent measurements of these quantities using optical broad band photometry. Additionally, our best fit model implies an average dust temperature of $\sim 40\: \rm{K}$.

\item We measure a stronger than average radiation field near our sources 7 and field C. This result is consistent with the sequential star formation scenario adopted in \citet{Tosaki07}, according to which a second-generation of star formation in the location of these sources has been triggered by the mechanical input from the first generation, 4~Myr old main cluster. 
 
\item We find a positive correlation between the strength of the 17~\mum\ PAH feature, the enhancement of the [\silii]/[\neii] emission, and the strength of the X-ray field towards our sources. We propose that X-rays are responsible for both the excitation of [\silii] and the enhancement of the 17~\mum\ feature via selective destruction of small PAH molecules.

\item Our detection of molecular hydrogen in the region indicates gas excitation temperatures of $\sim 500\: \rm{K}$ near selected sources, and a slightly larger abundance of $\rm{H}_2$ near field A, which correlates well with the location of a bright CO cloud reported in \citet{Wilson92}.

\end{enumerate}

\acknowledgements
The authors want to thank M\'ed\'eric Boquien and the HERM33ES team for kindly providing the PACS photometry of NGC~604, Dr.~Inga Kamp for her comments on the text, and the anonymous referee for a very detailed report. This paper would not have been possible without the hospitality of all the astronomical community at the Lowell Observatory. This research has been possible thanks to funds provided by the Netherlands Research School of Astronomy (NOVA). We made use of SAOImage DS9, developed by Smithsonian Astrophysical Observatory.

\clearpage

\end{document}